\documentclass[3p]{elsarticle}
\usepackage[english]{babel}
\usepackage{amsmath}
\usepackage{empheq,etoolbox}
\usepackage{amsfonts}
\usepackage{amssymb}
\usepackage{amsthm}
\usepackage{siunitx}
\usepackage{graphicx}
\usepackage{array}
\usepackage{tabularx}
\usepackage{rotating}
\usepackage{multicol}
\usepackage{makecell, cellspace, caption}
\usepackage[caption=false]{subfig}
\usepackage{multirow}
\usepackage{pifont}
\usepackage[table, svgnames, dvipsnames]{xcolor}
\usepackage{tikz}
\usetikzlibrary{fit}
\usepackage{booktabs}
\usepackage{ctable}
\usepackage{mathtools}
\usepackage{fancyhdr}
\usepackage[colorlinks = true,
            linkcolor = blue,
            urlcolor  = blue,
            citecolor = blue,
            anchorcolor = blue]{hyperref}

\newcolumntype{Y}[1]{>{\centering\arraybackslash}p{#1}}

\DeclarePairedDelimiter\norm{\lVert}{\rVert}%
\newcolumntype{Y}[1]{>{\centering\arraybackslash}p{#1}}
\newcommand\Tstrut{\rule{0pt}{2.6ex}}         
\newcommand\Bstrut{\rule[-0.9ex]{0pt}{0pt}}   

\usetikzlibrary{arrows,automata,backgrounds,fit,decorations.pathreplacing}

\let\today\relax
\makeatletter
\def\ps@pprintTitle{%
	\let\@oddhead\@empty
	\let\@evenhead\@empty
	\def\@oddfoot{\footnotesize\itshape
		{\today} \hfill}%
	\let\@evenfoot\@oddfoot
}
\makeatother

\journal{Name of the Journal} 

\begin{document}
	
\begin{frontmatter}

\title{Reacting condensed phase explosives in direct contact.} 

\author[CAMBRIDGE]{R. Dematt\`e}
\ead{rd609@cam.ac.uk}
\address[CAMBRIDGE]{Laboratory
for Scientific Computing, Cavendish Laboratory, Department of Physics, University of Cambridge, UK}
\author[CAMBRIDGE]{L. Michael}
\ead{lm355@cam.ac.uk}
\author[CAMBRIDGE]{N. Nikiforakis}
\ead{nn10005@cam.ac.uk}

\begin{abstract}

In this article we present a new formulation and an associated algorithm for the simultaneous numerical simulation of multiple condensed phase explosives in direct contact with each other, which may also be confined by (or interacting with one or more) compliant inert materials. Examples include composite rate-stick (i.e.\ involving two explosives in contact) problems and interaction of shock waves with chemically-active particles in condensed-phase explosives. There are several formulations which address the compliant or structural response of confiners and particles due to detonations, but the direct interaction of explosives remains a challenge for most formulations and algorithms. The proposed formulation addresses this problem by extending the conservation laws and mixture rules of an existing hybrid formulation (suitable for solving problems involving the coexistence of reactants and products in an explosive mixture and its immiscible interaction with inert materials) to model the interaction of multiple explosive mixtures. 
An algorithm for the solution of the resulting system of partial differential equations is presented, which includes a new robust method for the retrieval of the densities of the constituents of each explosive mixture. This is achieved by means of a multi-dimensional root-finding algorithm which employs physical as well as mathematical considerations in order to converge to the correct solution. The algorithm is implemented in a hierarchical adaptive mesh refinement framework and validated against results from problems with known solutions. It is evaluated for robustness for rate-stick and shock-induced flows in particle-laden explosives case-studies. It is shown that the method can simulate the interaction of detonation waves produced by military grade and commercial explosives in direct contact, each with its own distinct equation of state and reaction rate law. The ability of the new model to simulate reactive particles which are explicitly resolved in a heterogeneous explosive is demonstrated by a case-study of a shock wave interacting with a high explosive bead embedded in liquid nitromethane.\end{abstract}
\end{frontmatter}

\section{Introduction}

This paper is concerned with the numerical simulation of ignition and transition to detonation of condensed-phase commercial- and military-grade explosives. Of particular interest is the case of two distinct explosives which are in direct contact (i.e.\ separated by a material interface), but each is composed of two or more miscible components. This situation arises in several applications including the shock-induced ignition of explosive mixtures laden with reactive beads and the study of devices (such as detonators) which have several explosives in direct contact. Although there exist several formulations (see for instance Saurel and Abgrall \cite{SaurelAbgrall_1999}, Schoch et al. \cite{Schoch_2013} and Banks et al. \cite{Banks_Kapila}) which address the compliant or structural response of inert confiners and non-reacting particles due to detonations, these are limited to a single explosive material with one or more inert neighbours as the robust and accurate simulation of direct interaction of reacting interfaces remains a challenge for most algorithms, especially if there is a large density gradient across them and/or there is a co-existence of miscible and immiscible materials. The problem is further compounded if the mathematical model employs realistic equations of state and complex reaction rate laws.

Level-set based algorithms have been employed in the past to address 
the challenges of more than one reacting explosives in direct contact by Ioannou and Nikiforakis \cite{IoannouI:2020,IoannouII:2020}, in the context of resolved simulations of detonators, and by Kim et al.\cite{Kim2019}, who  investigated the enhanced blast effect generated by the combustion of metal particles dispersed in a high explosive. 
Diffuse interface approaches have been extensively used for the numerical simulation of explosives and their interaction with confiners. These may be broadly be divided into (i) models that are based on an augmented version of the Euler equations and (ii) models that are based on a multi-phase approach. 

Formulations of the augmented Euler type handle multi-material flows by using the Euler set of governing equations to describe the overall behaviour of the multi-component system ---which is assumed to be in mechanical and often also in thermal equilibrium--- and include additional evolution equations to describe the distinct constituents. Examples include the two- and three-material fluid-mixture model proposed by Banks et al. \cite{Banks_Schwendeman, Banks_Kapila}. Formulations based on a multi-phase approach, on the other hand, regard each component of the flow as a "phase" (e.g Saurel and Abgrall \cite{Saurel:1999a}, Saurel et al. \cite{Saurel2009}, Chinnayya et al. \cite{Chinnayya2004}). Each component has its own mass continuity equation and, depending on the closure assumptions, may also have its own momentum and energy equation. Models belonging to this class are in general full or reduced versions of the Baer-Nunziato system \cite{Baer:1986a}. Versions of the Baer-Nunziato system which are based on systematic asymptotic reductions in the limit of zero relaxation times have also been consider, including the five-equation (single-pressure and single-velocity) models of Kapila et al. \cite{Kapila_Menikoff}, Allaire et al. \cite{Allaire2002} and Murrone and Guillard \cite{MurroneGuillard}. 

Each of those approaches have advantages and disadvantages, often as a trade-off between accuracy, robustness and complexity. A formulation which combines the advantages of augmented Euler and multi-phase formulations and can simulate the interaction between miscible and immiscible components was proposed by Michael and Nikiforakis \cite{MINI16} (hereafter referred to as MiNi16). The original model offered a robust framework for modelling three-component flows which include two materials that are expected to physically mix (the explosive reactant and its products), and an interface between this mixture and a third inert material. Depending on which terms of the underlying partial differential equations are employed, reduced versions of MiNi16 can be employed for standalone simulations or in conjunction with formulations for computational fluid dynamics, elastoplastic materials or plasma, for arbitrary combinations of four states of matter \cite{Michael_Nikiforakis:2019_a,Michael_Nikiforakis:2019_b,Michael_Nikiforakis:2018,Michael_Millmore_Nikiforakis:2019}. 

However, the underlying mathematical formulation and numerical discretisation of MiNi16 is not valid for the case of two immiscible reactive materials interacting with each other. In this work, the mathematical model and the corresponding numerical algorithm of MiNi16 is extended by a suitable modification of the conservation laws and the corresponding changes in the mixture rules and sound speed calculations. A key element of the numerical solution is the robustness of the multi-dimensional root-finding method employed for the retrieval of the densities of the constituents of each explosive mixture. In order to  ensure convergence to the correct values, an algorithm which is solely based on the mathematics of the system is not sufficient, so the physics (thermodynamic properties) of the problem have to be taken into consideration. To this end, in this work we propose a methodology which provides an accurate initial guess as well as a suitable search domain as the basis for one- and two-dimensional root-finding algorithms. The complete model was validated for problems involving the interaction of ideal and non-ideal explosives. 

In the rest of this paper, the MiNi16 formulation is summarised in the first instance. The extended mathematical model is presented in section \ref{Sect:MINI16}. The numerical methodology and the numerical issues related to the robust root-finding algorithms are presented in section \ref{Sect:Num_methodology}. The resulting algorithm is validated against exact and known numerical solutions in section \ref{Sect:Num_results}. It is shown that the new model is able to robustly simulate the direct interaction of reacting military-grade and commercial explosives, while retaining the properties of the original formulation.

\subsection{Summary of the original MiNi16 formulation}\label{Sect:MINI16}

The original MiNi16 formulation is summarised here, to form a basis for the extended mathematical model presented in the following section. The model is a combination of a fluid-mixture approach to account for the heterogeneous mixture of the explosive reactants and products, and a two-phase model to solve for the immiscible interaction between the explosive mixture and the inert material.
The composition of the total mixture is determined by the colour function $z_1 \in [0,1]$ which represents the volume fraction of the inert material (denoted as $material \ 1$) with respect to the volume of the global mixture. Equivalently, the quantity $z_2 = 1-z_1$ represents the volume fraction of the explosive material ($material \ 2$), which is in turn composed of the reactants ($material \ \alpha$), with properties $(\rho_\alpha, {\bf{u}}_\alpha, p_\alpha)$,
and the products ($material \ \beta$), with properties $(\rho_\beta, {\bf{u}}_\beta, p_\beta)$. Pressure and velocity equilibrium conditions are assumed between all three constituents:
\begin{equation}
\begin{cases}
p_2 = p_\alpha = p_\beta = p_1 = p  \;, \\
{\bf{u}}_2 = {\bf{u}}_\alpha = {\bf{u}}_\beta = {\bf{u}}_1  = {\bf{u}} \;,\end{cases}
\end{equation}
while temperature equilibrium is only invoked between the reactants and the products of the reactive mixture: $T_\alpha = T_\beta$. The mathematical model is defined by the following system of governing equations:
\small
\begin{subequations} \label{eqn:MiNi16System} 
  \begin{empheq}[left=\empheqlbrace]{align}
  & \displaystyle{ \frac{\partial z_1}{\partial t} + {\bf{u}} \cdot \nabla z_1  = 0} \;, \\
  & \displaystyle{\frac{\partial z_1\rho_1}{\partial t} + \nabla \cdot(z_1 \rho_1 {\bf{u}}) = 0 }  \;, \\ 
  & \displaystyle{\frac{\partial z_2\rho_2}{\partial t} + \nabla \cdot(z_2 \rho_2 {\bf{u}}) = 0 } \;, \\
  & \displaystyle{\frac{\partial \rho {\bf{u}} }{\partial t} + \nabla \cdot(\rho {\bf{u}} \otimes {\bf{u}} + p \mathbb{I} ) = 0 }  \;, \\ 
  & \displaystyle{\frac{\partial E }{\partial t} + \nabla \cdot({\bf{u}}( E + p)) = 0 } \;, \\
  & \displaystyle{\frac{\partial z_2 \rho_2 \lambda }{\partial t} + \nabla \cdot(z_2 \rho_2 \lambda {\bf{u}}) = z_2\rho_2 \mathcal{K} } \;,
\end{empheq}
\end{subequations}
\normalsize
where $\rho$, ${\bf{u}}$ and $p$ are the density, velocity and pressure of the global mixture. The total energy $E$ is given by $E = \rho e + 1/2 \rho \norm{{\bf{u}}}^2$, where $e$ is the total specific internal energy.
System \eqref{eqn:MiNi16System} features two distinct continuity equations ({\ref{eqn:MiNi16System}b-c}), representing the conservation of mass for $material \ 1$  and $material \ 2$, as well as conservation laws ({\ref{eqn:MiNi16System}d-e}) for the momentum and total energy of the three-material mixture. Note that the colour function $z_1$ is governed by a non-conservative advection equation ({\ref{eqn:MiNi16System}a}); the quantity $\lambda$ is the mass fraction of the reactants ($material \ \alpha$) with respect to the explosive mixture and it is also governed by a simple advection equation with a source term $K_\lambda$ modelling the rate of conversion of reactants to products. This equation is then combined with the continuity equation for $material \ 2$ and written in conservative form ({\ref{eqn:MiNi16System}f}). The mixture rules used for the density and internal energy of the three-material mixture are:
\begin{subequations}\label{eqn:MINI16MixRules} 
  \begin{empheq}[left=\empheqlbrace]{align}
  & \displaystyle{ \rho = z_1\rho_1 + z_2\rho_2} \;, \hspace{0.5cm} \displaystyle{\frac{1}{\rho_2} = \frac{\lambda}{\rho_{\alpha}} +  \frac{1-\lambda}{\rho_{\beta}}}  \;; \\ 
  & \displaystyle{\rho e = z_1\rho_1 e_1 + z_2\rho_2 (\lambda e_{\alpha} + (1-\lambda)e_{\beta}) }  \;.
\end{empheq}
\end{subequations}
These relate the density and internal energy of the global mixture to the corresponding quantities of the individual materials.

\subsection{Equations of state (EoS)}

In this work, we assume that all materials are governed by an equation of state in the general Mie-Gr{\"u}neisen form
\begin{equation}\label{eqn:Mie_Grun}
p = p_{ref}(\rho) + \Gamma \rho( e - e_{ref}(\rho)) \;,
\end{equation}
where $p_{ref}(\rho)$ and $e_{ref}(\rho)$ are the reference pressure and reference energy, respectively; $\Gamma$ is the Gr{\"u}neisen coefficient, defined as 
\begin{equation}\label{eqn:Grun_Gamma}
\left. \Gamma = \frac{1}{\rho} \frac{\partial p}{\partial e} \right|_{\mathrlap{\rho}} \;.
\end{equation}
Depending on the particular form of the reference functions, different types of equations of state can be retrieved. Representative examples are:
\begin{itemize}[leftmargin=*]
\item[•] {\it{The Jones-Wilkins-Lee (JWL) EoS}} \cite{JWL_1968}, which is extensively employed to model reaction products and sometimes also the reactants. The JWL EoS uses isentropic reference curves with reference pressure and energy given by:
{\[\setlength{\multlinegap}{0pt}
\left\{
\begin{multlined}\label{eqn:JWL_RefCurves}
p_{REF}(\rho) = \mathcal{A}\exp(-\mathcal{R}_1 \frac{\rho_0}{\rho}) + \mathcal{B}\exp(-\mathcal{R}_2\frac{\rho_0}{\rho}) \;, \hspace{2.6cm} \\
e_{REF}(\rho) = \dfrac{\mathcal{A}}{\mathcal{R}_1\rho_0}\exp(-\mathcal{R}_1\frac{\rho_0}{\rho}) + \dfrac{\mathcal{B}}{\mathcal{R}_2\rho_0}\exp(-\mathcal{R}_2\frac{\rho_0}{\rho})\;, \hspace{4.3cm}
\end{multlined} \right.\]}where $\rho_0$ is the initial density and $\mathcal{A} \SI{}{\left[\pascal\right]}$, $\mathcal{B} \SI{}{\left[\pascal\right]}$, $\mathcal{R}_1 \SI{}{\left[ - \right]}$, $\mathcal{R}_2 \SI{}{\left[ - \right]}$ are empirically fitted parameters. A constant value for the Gr{\"u}neisen coefficient $\Gamma(\rho) = \Gamma_0$ is assumed. 
\item[•] {\it{The Cochran-Chan EoS}}. Condensed phase explosives (solid or liquid) can be also modelled using the Cochran-Chan EoS \cite{cochran_chan}, with reference pressure and energy given by
{\[\setlength{\multlinegap}{0pt}
\left\{
\begin{multlined}\label{eqn:CC_RefCurves}
p_{REF}(\rho) = \mathcal{A}\left(\ \dfrac{\rho_0}{\rho} \right)^{-\mathcal{E}_1} - \mathcal{B}\left(\ \dfrac{\rho_0}{\rho} \right)^{-\mathcal{E}_2}\;, \\
e_{REF}(\rho) = \dfrac{\mathcal{-A}}{\rho_0(1-\mathcal{E}_1)}\left[\ \left(\dfrac{\rho_0}{\rho} \right)^{-\mathcal{E}_1} -1 \right] + \\ \dfrac{-\mathcal{B}}{\rho_0(1-\mathcal{E}_2)}\left[\ \left(\dfrac{\rho_0}{\rho} \right)^{-\mathcal{E}_2} -1 \right] \;,
\end{multlined}
\right.\]}and with a constant Gr{\"u}neisen coefficient; here again $\rho_0$ is the initial density and the parameters $\mathcal{A} \SI{}{\left[\pascal\right]}$, $\mathcal{B}\SI{}{\left[\pascal\right]}$, $\mathcal{E}_1\SI{}{\left[-\right]}$, $\mathcal{E}_2\SI{}{\left[-\right]}$ are calibrated using experimental data for the material under consideration. 
\end{itemize}
In reacting flows, a term $Q \left[\SI{}{\meter^2\second^{-2}}\right]$ is included in the reference energy function of the products, representing the heat of
detonation released during the reactions, such that $\tilde{e}^{p}_{REF}(\rho) = {e}^{p}_{REF}(\rho) - Q $. It can be easily verified that this is equivalent to having an inhomogeneous evolution equation for the total energy of the system.

\subsubsection{Recovery of temperature}\label{Sect:Temperature_Recovery}

Computing the temperature of a condensed phase explosive in general involves extending the mechanical EoS model of the form $p=p(e,v)$ to a complete form that is consistent with the first law of thermodynamics: $\mathrm{d}e = T\mathrm{d}s - p \mathrm{d}v$ (where $v$ is the specific volume and $s$ is the specific entropy). For equations of state in the general Mie-Gr{\"u}neisen form \eqref{eqn:Mie_Grun}, this is typically accomplished by supplementing the EoS with an additional reference function for temperature, $T_{REF}(\rho)$, analogous to the reference functions for pressure and energy. When the reference curve is an isentrope and a constant value for the Gr{\"u}neisen gamma is assumed (as in the case of the JWL EoS), the reference temperature can be easily computed as: $T_{REF}(\rho) = T_0 \left(\frac{\rho}{\rho_0} \right)^{\Gamma_0}$. Then, given any state $\left(e, \rho \right)$ off the reference curve, the specific heat capacity at constant volume $c_v = \left. \frac{\partial e}{\partial T} \right|_{\mathrlap{v}}$ (here assumed to be constant $c_v(\rho, T) = \bar{c_v}$), can be used to determine the corresponding temperature $T(e,\rho)$:
\begin{equation}\label{eqn:T_off_curve}
T(e,\rho) - T_{REF}(\rho) = \frac{e-e_{REF}(\rho)}{\bar{c_v}} = \frac{p-p_{REF}(\rho)}{\rho\Gamma\bar{c_v}} \;.
\end{equation}

\subsection{Reaction rate models}

The modular structure of the algorithm does not restrict the choice of reaction rate law for the explosives. In this work, two reaction rate models are employed which are chosen so that comparisons of our results against published benchmarks can be performed. These are the Ignition and Growth ({\it{I\&G}}) model developed by Lee and Tarver \cite{Lee&Tarver}, and a single-step temperature-based Arrhenius rate law. These two models are briefly described below. Other models can be implemented, as necessary.

\subsubsection{The Ignition and Growth model}

In the {\it{I\&G}} model, the reaction mechanism is described as a multi-stage process involving different contributing terms, each of which represents a distinct physical phenomenon. In particular, the {\it{I\&G}} reaction rate law is given by
\begin{equation}\label{eqn:IG_model1}
\dfrac{\mathrm{D}\Phi}{\mathrm{D}t} = - \dfrac{\mathrm{D}\lambda}{\mathrm{D}t} = - \mathcal{K} = \mathcal{R}_I+\mathcal{R}_{G_1}+\mathcal{R}_{G_2} \;,
\end{equation}
and the three terms are defined as
\begin{equation}\label{eqn:IG_model2}
\begin{aligned}
\mathcal{R}_I &= I(1-\Phi)^b(\rho/\rho_0 -1 - a)^x \mathcal{H}(\Phi_{ig,max}-\Phi) \;, \\
\mathcal{R}_{G_1} &= \tilde{G_1}(1-\Phi)^c\Phi^dp^y\mathcal{H}(\Phi_{G_1,max}-\Phi) \;, \\
\mathcal{R}_{G_2} &= \tilde{G_2}(1-\Phi)^e\Phi^gp^z\mathcal{H}(\Phi-\Phi_{G_2,max}) \;,
\end{aligned}
\end{equation}
where $\Phi = 1-\lambda$ is the mass fraction of the products, $\mathcal{H}(x)$ is the Heaviside function and $p$ and $\rho$ are the pressure and density of the explosive respectively. The constants $I \SI{}{\left[\second^{-1}\right]}$, $a$, $b$, $c$, $d$, $e$, $g$, $x$, $y$, $z$, $\tilde{G_1} \SI{}{\left[\pascal^{-y}\second^{-1}\right]}$ and $\tilde{G_2} \SI{}{\left[\pascal^{-z}\second^{-1}\right]}$ are calibrated for each particular explosive as are $\Phi_{ig,max}$, $\Phi_{G_1,max}$, $\Phi_{G_2,max}$ which limit the contribution of each term to a specific stage of the reaction progress. The form of this reaction rate was constructed so that each term represents one of the three stages of reaction observed during the shock initiation and detonation of pressed solid explosives. For a more detailed interpretation of the three terms the reader is referred to Tarver \cite{Tarver2005}.

\subsubsection{Temperature-dependent Arrhenius reaction rate law}\label{Sect:Arrhenius_law}

A single-step temperature-based Arrhenius rate law is given by
\begin{equation}\label{eqn:Arrhenius_law}
\dfrac{\mathrm{D}\lambda}{\mathrm{D}t} = \mathcal{K} = -\lambda A \exp{(-\frac{T}{T_A})} \;,
\end{equation}
where $A \SI{}{\left[\second^{-1}\right]}$ is a pre-exponential factor, $T$ is the temperature of the reactive mixture and $T_A \SI{}{\left[\kelvin\right]}$ is the activation temperature. 

\section{Extended MiNi16 formulation}\label{Sect:Extend Formulation}

\begin{figure}
  \centering
    \includegraphics[width=0.475\textwidth]{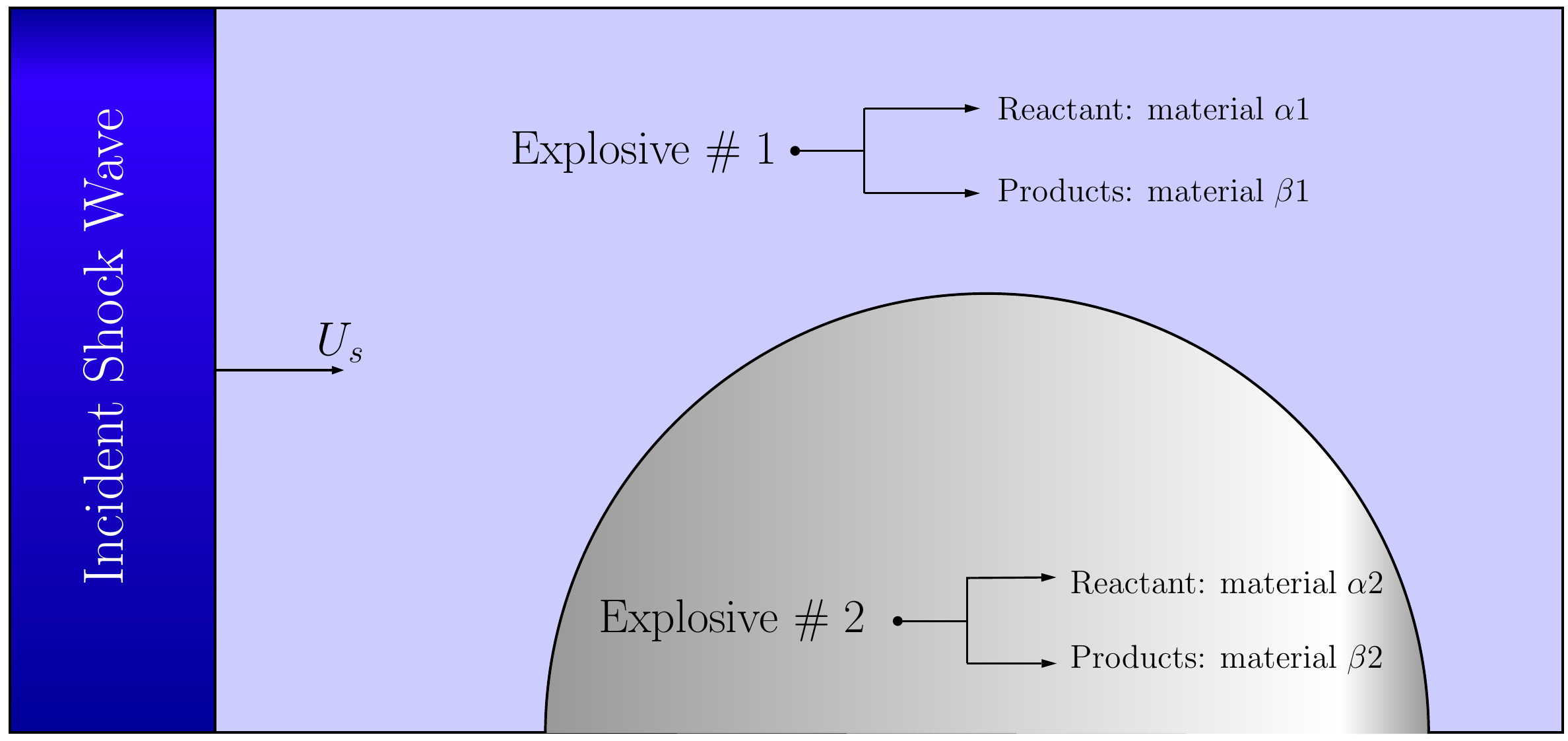} 
  \caption{\label{Fig:Formulation_Schematic}Schematic of a possible application of interest for this work. The four materials as considered by the extended formulation proposed in this work are illustrated.}
\end{figure}

In order to describe the situation where two distinct explosives (i.e.\ two sets of miscible materials) are in direct contact with each other, see for example Fig.~\ref{Fig:Formulation_Schematic}, the original model has been modified by incorporating a second augmented-Euler-type fluid-mixture model \cite{Banks_Schwendeman}, meaning that each phase of the original five-equation interface model \cite{Allaire2002} is, in this case, a two-material fluid-mixture system. As a result, the proposed extended formulation can take two explosive phases, each of which can be a reactant-product mixture. Let us consider the situation shown in Fig.~\ref{Fig:Formulation_Schematic}, where the two explosive phases are separated by a material interface and no physical mixing takes place  between them. In this case, each reactive phase ($i=1,2$) is in turn composed of two components: the explosive reactants (material $\alpha_i$), with properties $(\rho_{\alpha_i}, p_{\alpha_i}, u_{\alpha_i})$ and the explosive products (material $\beta_i$), with properties $(\rho_{\beta_i}, p_{\beta_i}, u_{\beta_i})$. The mass fraction of the reactants  with respect to the corresponding explosive mixture, which has density $\rho_i$, is denoted by $\lambda_i$; as a result, the mass fraction of the products with respect to the first explosive mixture is, trivially, $1-\lambda_i$.
In the spirit of the original MiNi16 formulation, 
pressure and velocity equilibrium conditions are assumed between all materials:
\begin{equation}
\begin{cases}
p_1 = p_{\alpha_1} = p_{\beta_1} = p_2 = p_{\alpha_2} = p_{\beta_2} = p  \;, \\
{\bf{u}}_1 = {\bf{u}}_{\alpha_1} = {\bf{u}}_{\beta_1} = {\bf{u}}_2 = {\bf{u}}_{\alpha_2} = {\bf{u}}_{\beta_2}  = {\bf{u}} \;,\end{cases}
\end{equation}
while an isothermal closure condition is only assumed between the reactants and the products of each explosive mixture, i.e.\ $T_{\alpha_i} = T_{\beta_i} = T_i$, for $i=1,2$. Note that other closure conditions between the constituents of each reactive phase may be more suitable,  depending on the time scales of the problem under consideration, see for instance Stewart et al. \cite{Stewart2002}, and could also be implemented. The governing equations for this system are given by
\begin{subequations}\label{eqn:SystemExtended}  
\begin{empheq}[left=\empheqlbrace]{align}
  & \displaystyle{ \frac{\partial z_1}{\partial t} + {\bf{u}} \cdot \nabla z_1  = 0} \;, \\
  & \displaystyle{\frac{\partial z_1\rho_1}{\partial t} + \nabla \cdot(z_1 \rho_1 {\bf{u}}) = 0 }  \;, \\ 
  & \displaystyle{\frac{\partial z_2\rho_2}{\partial t} + \nabla \cdot(z_2 \rho_2 {\bf{u}}) = 0 } \;, \\
  & \displaystyle{\frac{\partial \rho {\bf{u}} }{\partial t} + \nabla \cdot(\rho {\bf{u}} \otimes {\bf{u}} + p \mathbb{I} ) = 0 }  \;, \\ 
  & \displaystyle{\frac{\partial E }{\partial t} + \nabla \cdot({\bf{u}}( E + p)) = 0  } \;, \\
  & \displaystyle{\frac{\partial z_1 \rho_1 \lambda_1 }{\partial t} + \nabla \cdot(z_1 \rho_1 \lambda_1 {\bf{u}}) = z_1\rho_1 \mathcal{K}_1 }  \;, \\ 
  & \displaystyle{\frac{\partial z_2 \rho_2 \lambda_2 }{\partial t} + \nabla \cdot(z_2 \rho_2 \lambda_2 {\bf{u}}) = z_2\rho_2 \mathcal{K}_2 } \;.
\end{empheq}
\end{subequations}
Note that system \eqref{eqn:SystemExtended} is identical to \eqref{eqn:MiNi16System}, but with two distinct transport equations which separately evolve the reactant mass fraction variables, $\lambda_1$ and $\lambda_2$.

\subsection{Mixture rules and numerical evaluation of the mixed equation of state}\label{Sect:MixRulesandMixedEoS}

To complete the proposed formulation and ensure the appropriate numerical mixing between the different materials, the definition of a set of mixture rules is required. These are derived by naturally extending the approach used by Michael and Nikiforakis \cite{MINI16}, 
wherein multi-phase type and fluid-mixture type mixture rules have been opportunely combined.

The mixture rule for the internal energy is:
\begin{subequations}\label{eqn:MixtureRuleEnergy}
  \begin{empheq}[left=\empheqlbrace]{align}
  & \displaystyle{ \rho e = z_1\rho_1 e_1 + z_2\rho_2 e_2} \;, \\
  & \displaystyle{e_i = \lambda_i e_{\alpha_i} + (1-\lambda_i)e_{\beta_i}, \ \ \rm{for} \ i=1,2 } \;.
  \end{empheq} 
\end{subequations}
Similarly, by volume averaging, the density of the overall mixture $\rho$ can be expressed in terms of the phase densities $\rho_1$ and $\rho_2$ by the weighted average
\begin{equation}
\rho = z_1\rho_1 + z_2\rho_2 \;.
\end{equation}
The density of each explosive mixture ($i=1,2$) can be in turn related to the densities of its constituents as 
\begin{equation}\label{eqn:MixRulesDensityExt}
\dfrac{1}{\rho_i} = \dfrac{\lambda_i}{\rho_{\alpha_i}} + \dfrac{1-\lambda_i}{\rho_{\beta_i}}
\end{equation}

Finally, as in the original MiNi16 model \cite{MINI16}, a mixture rule for the inverse Gr{\"u}neisen coefficient of the global mixture is defined as 
\begin{equation}\label{eqn:MixtureRuleInvGrun}
\xi = \dfrac{1}{\Gamma} = z_1\xi_1 + z_2\xi_2 \;,
\end{equation}
with $\xi_i = \left. \frac{\partial \rho_i e_i}{\partial p} \right|_{\mathrlap{\rho_i}}$. By substitution of the equations of state of each individual component into \eqref{eqn:MixtureRuleEnergy} and application of the pressure equilibrium condition between all constituents, a mixture equation of state is then defined, which is required to compute the pressure $p$; that is:
\begin{eqnarray}\label{eqn:MixedEOS_energy}
\rho e = &&p\sum\limits_{i}(\dfrac{z_i\rho_i \lambda_i}{\rho_{\alpha_i}\Gamma_{\alpha i }} + \dfrac{z_i\rho_i(1-\lambda_i)}{\rho_{\beta_i}\Gamma_{\beta_i}})+ \nonumber\\ &&\sum\limits_{i} z_i\rho_i\Big(\lambda_i {\rm{REF}}_{\alpha_i } + (1-\lambda_i) {\rm{REF}}_{\beta_i}\Big) \;,
\end{eqnarray}
with 
\begin{equation}
{\rm{REF}}_k = e_{ref_k}(\rho_k) - \frac{p_{ref_k}(\rho_k)}{\rho_k\Gamma_k} \hspace{0.25cm} {\rm{for}} \hspace{0.2cm} k = \alpha_1, \alpha_2, \beta_1, \beta_2 \;.
\end{equation}
Rearranging equation \eqref{eqn:MixedEOS_energy} for $p$ gives:
\begin{equation}\label{eqn:MixedEOS_pressure}
p = \dfrac{1}{\sum\limits_{i}(\dfrac{z_i\rho_i \lambda_i}{\rho_{\alpha_i}\Gamma_{\alpha_i }} + \dfrac{z_i\rho_i(1-\lambda_i)}{\rho_{\beta_i}\Gamma_{\beta_i}})}\Big[\rho e - \sum\limits_{i} z_i\rho_i\Big(\lambda_i {\rm{REF}}_{\alpha_i } + (1-\lambda_i) {\rm{REF}}_{\beta_i}\Big)\Big] \;.
\end{equation}

It is important to note that the individual component densities $\rho_k$ (for $k = \alpha_1, \alpha_2, \beta_1, \beta_2$) in \eqref{eqn:MixedEOS_pressure} are unknown, as they are not evolved by the system of governing equations \eqref{eqn:SystemExtended}. Under the assumption of temperature equilibrium between the constituents of each explosive mixture, i.e.\ $T_{\alpha_i} = T_{\beta_i} = T_i$ (for $i=1,2$), the unknown individual component densities are retrieved through a robust two-dimensional root-finding algorithm. This is presented in full detail in Sect.~\ref{Sect:Root Finding problem}.

\subsection{Calculation of sound speed and the hyperbolicity of the system}

Following the derivation given by Allaire et al. \cite{Allaire2002}, the sound speed of the global mixture is computed as
\begin{equation}\label{eqn:MixSoundspeed}
c^2 = \dfrac{Y_1 \xi_1 c_1^2 + Y_2\xi_2 c_2^2}{\xi} \;,
\end{equation}
where $Y_i = \frac{z_i\rho_i}{\rho}$ represents the phase mass fraction and $\xi$ is given by \eqref{eqn:MixtureRuleInvGrun}. In order to compute the sound speed
\begin{equation}\label{eqn:sound_speed_c12}
c_i^2 = \dfrac{\dfrac{p}{\rho_i^2} - \left. \dfrac{\partial e_i}{\partial \rho_i} \right|_{p, \lambda_i}}{\left. \dfrac{\partial e_i}{\partial p} \right|_{\rho_i, \lambda_i}}
\end{equation}
and the inverse Gr{\"u}neisen coefficient $\xi_i = \rho_i \left. \frac{\partial e_i}{\partial p} \right|_{\rho_i,\lambda_i}$ of each explosive phase ($i=1,2$),
the calculation of mixture quantities $\left. \frac{\partial e_i}{\partial \rho_i} \right|_{p, \lambda_i}$ and $\left. \frac{\partial e_i}{\partial p} \right|_{\rho_i, \lambda_i}$ is required. With the use of implicit differentiation, as in Ioannou \cite{Ioannou_PhD}, these are found to be
\begin{equation}\label{eqn:Sound_partial1_mat1}
\left. \dfrac{\partial e_i}{\partial \rho_i} \right|_{\mathrlap{p, \lambda_i}} = \dfrac{\rho_{\alpha_i}^2\rho_{\beta_i}^2\left(\lambda_i \left. \dfrac{\partial e_{\alpha_i}}{\partial \rho_{\alpha_i}} \right|_{\mathrlap{p}} \ \ \left. \dfrac{\partial T_{\beta_i}}{\partial \rho_{\beta_i}} \right|_{\mathrlap{p}} + (1-\lambda_i)\left. \dfrac{\partial e_{\beta_i}}{\partial \rho_{\beta_i}} \right|_{\mathrlap{p}} \ \ \left. \dfrac{\partial T_{\alpha_i}}{\partial \rho_{\alpha_i}} \right|_{\mathrlap{p}} \ \right)}{\rho_i^2\left(\lambda_i\rho_{\beta_i}^2\left. \dfrac{\partial T_{\beta_i}}{\partial \rho_{\beta_i}} \right|_{\mathrlap{p}} + (1-\lambda_i)\rho_{\alpha_i}^2\left. \dfrac{\partial T_{\alpha_i}}{\partial \rho_{\alpha_i}} \right|_{\mathrlap{p}} \ \right)} 
\end{equation}
\begin{equation}\label{eqn:Sound_partial1_mat12}
\left. \dfrac{\partial e_i}{\partial p} \right|_{\mathrlap{\rho_i, \lambda_i}} = \lambda_i\left. \dfrac{\partial e_{\alpha_i}}{\partial p} \right|_{\mathrlap{\rho_{\alpha_i}}} + (1-\lambda_i)\left. \dfrac{\partial e_{\beta_i}}{\partial p} \right|_{\mathrlap{\rho_{\beta_i}}} -\dfrac{\lambda_i(1-\lambda_i)\left(\rho_{\alpha_i}^2\left. \dfrac{\partial e_{\alpha_i}}{\partial \rho_{\alpha_i}} \right|_{\mathrlap{p}} -\rho_{\beta_i}^2\left. \dfrac{\partial e_{\beta_i}}{\partial \rho_{\beta_i}} \right|_{\mathrlap{p}} \ \ \right)\left(\left. \dfrac{\partial T_{\alpha_i}}{\partial p} \right|_{\mathrlap{\rho_{\alpha_i}}}- \left. \dfrac{\partial T_{\beta_i}}{\partial p} \right|_{\mathrlap{\rho_{\beta_i}}} \ \ \ \right) }{\lambda_i\rho_{\beta_i}^2\left. \dfrac{\partial T_{\beta_i}}{\partial \rho_{\beta_i}} \right|_{\mathrlap{p}} + (1-\lambda_i)\rho_{\alpha_i}^2\left. \dfrac{\partial T_{\alpha_i}}{\partial \rho_{\alpha_i}} \right|_{\mathrlap{p}}} \;. 
\end{equation}
Restricting the analysis to the one dimensional case, the system of equations \eqref{eqn:SystemExtended}, without source terms and for smooth solutions, can be written in quasi-linear form using the vector of primitive variables $\textbf{W} = \left(z_1, \rho_1, \rho_2, u, p, \lambda_1, \lambda_2 \right)^T $:
\begin{equation}
\partial_t{\bf{W}} + \underline{\underline{\mathcal{A}}}({\bf{W}})\partial_x{\bf{W}} = {\bf{0}} \;,
\end{equation}
where:
\begin{equation}
\underline{\underline{\mathcal{A}}}({\bf{W}}) =
\begingroup 
\setlength\arraycolsep{4.pt}
\begin{bmatrix}
u & 0 & 0 & 0 & 0 & 0 & 0 \\[0.1cm]
0 & u & 0 & \rho_1 & 0 & 0 & 0 \\[0.1cm]
0 & 0 & u & \rho_2 & 0 & 0 & 0 \\[0.1cm]
0 & 0 & 0 & u & \frac{1}{\rho} & 0 & 0 \\[0.1cm]
0 & 0 & 0 & \rho c^2 & u & 0 & 0 \\[0.1cm]
0 & 0 & 0 & 0 & 0 & u & 0 \\[0.1cm]
0 & 0 & 0 & 0 & 0 & 0 & u \\
\end{bmatrix} \;,
\vspace{0.1cm}
\endgroup
\end{equation}
It can be easily verified that the Jacobian matrix $\underline{\underline{\mathcal{A}}}({\bf{W}})$ has eigenvalues given by
\begin{equation}
\lambda_1 = u - c, \hspace{0.2cm} \lambda_2 = \lambda_3 = \lambda_4 = \lambda_5 = \lambda_6 = u, \hspace{0.2cm} \lambda_7 = u + c \;. 
\end{equation}
Hence, if the sound speed $c$ as given by \eqref{eqn:MixSoundspeed} is positive, $\underline{\underline{\mathcal{A}}}({\bf{W}})$ is diagonalisable with real eigenvalues and a complete set of eigenvectors, which proves the hyperbolicity of the system.

\section{Numerical Methodology}\label{Sect:Num_methodology}

The extended hydrodynamic model developed in Sect.~\ref{Sect:Extend Formulation} constitutes a non-linear hyperbolic system with source terms. In two space dimensions, this can be written in the form:
\begin{equation}\label{eqn:system}
\partial_t \textbf{U} + \partial_x \textbf{F}(\textbf{U}) + \partial_y \textbf{G}(\textbf{U}) = \textbf{H}(\textbf{U}) + \textbf{S}(\textbf{U})\;,
\end{equation} 
with 
\begin{equation}
{\textbf{U}} = 
\begin{pmatrix}
z_1,
z_1\rho_1,
z_2\rho_2,
\rho u,
\rho v,
E,
z_1\rho_1\lambda_1,
z_2\rho_2\lambda_2, 
\end{pmatrix}^T
\end{equation} and 
\begin{subequations}\label{eqn:U-F1} 
  \begin{empheq} {align}
  & \displaystyle{{\textbf{F}}(\textbf{U}) = 
\begin{pmatrix}
z_1 u \\
z_1\rho_1 u \\
z_2\rho_2 u \\
\rho u^2 + p \\
\rho u v \\
u(E+p) \\
z_1\rho_1\lambda_1 u \\
z_2\rho_2\lambda_2 u
\end{pmatrix}, \hspace{0.12cm}
{\textbf{G}}(\textbf{U}) = 
\begin{pmatrix}
z_1 v \\
z_1\rho_1 v \\
z_2\rho_2 v \\
\rho u v \\
\rho v^2 + p \\
v(E+p) \\
z_1\rho_1\lambda_1 v \\
z_2\rho_2\lambda_2 v
\end{pmatrix}}  \;, \\ 
  & \displaystyle{{\textbf{H}}(\textbf{U}) = 
\begin{pmatrix}
z_1 \nabla \cdot \textbf{u} \\
0 \\
0 \\
0 \\
0 \\
0 \\
z_1\rho_1\mathcal{K}_1 \\
z_2\rho_2\mathcal{K}_2
\end{pmatrix} , \hspace{0.12cm}
{\textbf{S}}(\textbf{U}) = -\frac{u_r}{r}
\begin{pmatrix}
0 \\
z_1\rho_1  \\
z_2\rho_2 \\
\rho u_r \\
\rho u_z \\
E+p \\
z_1\rho_1\lambda_1  \\
z_2\rho_2\lambda_2
\end{pmatrix}}  \;,
\end{empheq}
\end{subequations}
where ${\textbf{S}}(\textbf{U})$ is the geometric source term vector arising when axisymmetric problems are considered; in this case, $u_r$ and $u_z$ denote the radial and axial velocity components, respectively; $r$ is the radial distance from the axis of symmetry, which is considered to be at $r=0$. 

In order to simplify the process of solving the full system, second-order Strang operator splitting \cite{Strang:1968a} is used to advance the numerical solution in time. Over a sufficiently small time step $\Delta t$ (chosen according to the classical Courant-Friedrichs-Lewy (CFL) condition), this allows for the independent integration of the homogeneous part of the system through an appropriate hyperbolic solver and the utilisation of a dedicated ODE solver to evaluate the effect of the source terms. The hyperbolic part of the equations is solved using the HLLC solver \cite{Toro:1992e} in the conventional Godunov scheme \cite{Godunov:1959a}. To achieve second order of accuracy,  
the high-resolution MUSCL-Hancock approach \cite{vanLeer:1979a, Toro:2009a}, with the van Leer slope limiter \cite{vanLeer:1979a}, is used on the primitive variables. Moreover, Strang operator splitting is employed to account for all dimensions. 

It is important to note that the evolution equation of the volume fraction $z_1$
\begin{equation}\label{eqn:z1_NonCon}
\frac{\partial z_1}{\partial t} + {\bf{u}} \cdot \nabla z_1 = 0 
\end{equation}
is a non-conservative equation and as a result cannot be integrated as the rest of the system. Following Saurel et al. \cite{Saurel2009} and Michael and Nikiforakis \cite{MINI16}, a first-order update formula for $z_1$ can be obtained by straightforward application of the Godunov method for advection equations. That is, for the x-direction: 
\small
\begin{equation}\label{eqn:Non-Conserv_VolumeFraction}                       
z_{1i}^{n+1} = z_{1i}^{n} -\frac{\Delta x}{\Delta t}\left[ {(uz_1)}_{i+\frac{1}{2}}^{\star} - (uz_1)_{i-\frac{1}{2}}^{\star} - z_{1i}^{n}(u_{i+\frac{1}{2}}^{\star} - u_{i-\frac{1}{2}}^{\star} )\right] \;,
\end{equation}
\normalsize
where starred quantities represent quantities evaluated in the intermediate star region of the Riemann problem solution. 

The procedure for the remaining directions is entirely analogous, following Strang operator splitting again. A second-order accurate update formula for $z_1$ can be then constructed using the MUSCL approach, thus reconstructing and evolving the velocity $u$ and the volume fraction $z_1$ before inserting them in \eqref{eqn:Non-Conserv_VolumeFraction}. Finally, the reaction and geometric source terms are integrated using the implicit Gear algorithm \cite{CVODE}.

In numerical simulations of this work, the model is solved on a Cartesian mesh, and hierarchical adaptive mesh refinement \cite{Berger:1984a, Berger:1989a} is employed to better focus computational resources, increasing the resolution only across shock waves, detonation waves, material interfaces and other areas where variables exhibit complex dynamic behaviour. This is achieved by implementing our algorithms in the C++ AMReX platform \cite{AMReX2019}, which is an open source software library from Lawrence Berkeley National Laboratory. 
AMReX offers a parallel, hierarchical, block-structured adaptive mesh refinement (AMR) computational framework, which essentially represents a hierarchical inter-mesh communication scheme, based on block-structured mesh patches that locally and recursively increase the resolution of an underlying base coarse grid. Cells of identical resolution are grouped into logically rectangular sub-grids, with the coarse grid $G_0$ corresponding to $Level_0$ of the hierarchy and sub-grids within $G_0$ belonging to $Level_1$, and so on.  Refined grids are derived recursively from coarser ones, based upon a flagging criterion, which, in the present work is based on density gradients across cells. Subcycling in the discrete time steps is employed in order to bypass the global time step restriction imposed by the size of the smallest cells. For further details, the interested reader is referred to the website \href{https://amrex-codes.github.io/amrex/}{https://amrex-codes.github.io/amrex/}, where full documentation and source code are made available.

\subsection{Root-finding algorithm}\label{Sect:Root Finding problem}

As noted in Sect.~\ref{Sect:MixRulesandMixedEoS}, the individual constituent densities  $\rho_k$ (for $k = \alpha_1, \alpha_2, \beta_1, \beta_2$) are not explicitly evolved by the system of governing equations; their recovery, however, is necessary for evaluating the component equations of state, converting from conservative to primitive variables (and vice-versa) and computing the sound speed at each time step. To this end, temperature equilibrium between the reactants and the products of each reactive material is invoked and a two-dimensional root-finding procedure is then followed to evaluate the individual component densities. For clarity of exposition, in the next subsection we will first describe a robust one-dimensional root-finding technique developed in the context of the original MiNi16 formulation and then turn our attention to the more challenging two-dimensional root-finding method required by our extended model. We anticipate that the one-dimensional root-finding technique also plays a crucial role in the construction of the two-dimensional algorithm and for this reason is described in full detail. Also note that when presenting the one-dimensional root-finding method we assume a three-component system (as in the original MiNi16 model), where only phase $2$ is a reactant-product mixture.

\subsubsection{One-dimensional root-finding algorithm}\label{Sect:RootFinding1D}

The imposition of temperature equilibrium between the constituents of the explosive mixture corresponds to computing the root of the function $f_{T_{eq}}$ defined as 
\begin{equation}\label{eqn:Root1d_Teq}
f_{T_{eq}} = T_\alpha(p,\rho_\alpha) - T_\beta(p,\rho_\beta) \;.
\end{equation}
The temperatures are calculated using \eqref{eqn:T_off_curve} and the component densities ($\rho_\alpha$ and $\rho_\beta$) are strictly related through the mixture rule (\ref{eqn:MINI16MixRules}a). If the task at hand is to invert the mixed equation of state and compute the internal energy when the pressure is known, the root of the function $f_{T_{eq}}$ can be found directly. Equation \eqref{eqn:MixedEOS_energy} can be then used to compute the energy once the root has been found. If, instead, the pressure is unknown (but $\rho e$ is known) expression \eqref{eqn:MixedEOS_pressure} can be used to eliminate $p$ from the temperature equilibrium \eqref{eqn:Root1d_Teq}. 

When tackling the root-finding problem $f_{T_{eq}} = 0$ numerically, consideration must first be given to the choice of the independent variable. In particular, when the mixture is mainly composed of products ($\lambda < 0.5$), $\rho_\alpha$ is chosen as the independent variable and $\rho_\beta$ is obtained from (\ref{eqn:MINI16MixRules}a), while for mixtures with mainly reactants ($\lambda >0.5$) $\rho_\beta$ is taken as the independent variable and $\rho_\alpha$ is calculated accordingly. Note that this is necessary to have a well-conditioned problem when $\lambda$ approaches zero or one. The root of the function $f_{T_{eq}}$ is found by applying a modified version of the Dekker-Brent method \cite{Dekker:1969, Brent:1973}, where the Newton-Raphson method is used in place of the secant method. The algorithm can be summarised as follows:
\begin{itemize}[leftmargin=0.5cm]
\item[]{\bf{Step 1: Initial guess}}. The availability of a good starting value is of paramount importance for the fast convergence of the method to a physical solution. As suggested by Banks et al. \cite{Banks_Kapila}, the best choice for the starting value is represented by the converged solution at the previous time step.
\item[]{\bf{Step 2: Search domain}}. Then, it is fundamental to define a density range in which to search for the solution. This requires the determination of a lower and upper limit for the independent variable (hereafter simply denoted as $x$). The lower limit can be deduced from the requirement that the density of each material must be positive. When the independent variable is $\rho_\alpha$, this restriction can be expressed as: $\rho_\alpha > \lambda \rho_2$. If $\rho_\alpha$ violates this condition the corresponding value for $\rho_\beta = \frac{(1-\lambda)\rho_2\rho_\alpha}{\rho_\alpha -\lambda\rho_2}$ will be in fact negative. Analogously, when the independent variable is $\rho_\beta$, the density positivity requirement reads as: $\rho_\beta > (1-\lambda) \rho_2$.
Concerning the determination of the upper limit, one can take advantage of thermodynamic understanding of the objective function \cite{Wilkinson2017}. For example, we can observe that in the limit of large $\rho_\alpha$ we have $T_\alpha < T_\beta$ and therefore the function $f_{T_{eq}}$ will be large and negative. Similarly, in the limit of large $\rho_\beta$ we have $T_\alpha > T_\beta$ and $f_{T_{eq}}$ will be large and positive. On the basis of these considerations, the procedure used to set the upper limit is as follows:
\begin{itemize}[leftmargin=0.25cm]
\item[]{\it Case 1: $\rho_\alpha$ is the independent variable}. In the first stage we take $x_{max} = 2x^0$, where $x^0$ denotes the initial guess value. Then, we check whether the objective function evaluated at $x_{max} = 2x^0$ is negative. If this is not the case, we iteratively double $x_{max}$ until $f_{T_{eq}}$ is found to be negative.
\item[]{\it Case 2: $\rho_\beta$ is the independent variable}. The procedure is the same as above, but, in this case, we require $f_{T_{eq}}$ to be positive when evaluated at $x_{max}$.
\end{itemize}
\item[]{\bf{Step 3: Iterative procedure}}. Given an accurate initial guess and having established a suitable search domain, the iterative algorithm is now ready to start. At every iteration the objective function is evaluated, along with its derivative (an analytical expression can be found with the help of a symbolic manipulator), and an iteration of the Newton-Raphson method is carried out, that is
\begin{equation}
x^{(k+1)} = x^{(k)} - \dfrac{f_{T_{eq}}(x^{(k)})}{f_{T_{eq}}^{\prime}(x^{(k)})} \;.
\end{equation}
If the new value for the independent variable, $x^{(k+1)}$, falls within the search space, then the iteration is accepted and the search domain is updated on the basis of the considerations made above. If, instead, $x^{(k+1)}$ is not inside the search space, the Newton-Raphson iteration is rejected and the bisection method is applied:
\begin{equation}
x^{(k+1)} = \dfrac{x^{(k)}_{min} +x^{(k)}_{max}}{2} \;.
\end{equation}
The algorithm terminates when the standard stopping criterion of the Dekker-Brent method is reached.
\end{itemize}

A final remark on the method is that when the algorithm does not converge or converges to an unphysical solution, a different starting value is tried. Possible choices in this regard are:
\begin{itemize}[leftmargin=*]
\item[•] $\rho_\alpha^0 = \rho_\beta^0 = \rho_2$.
\item[•] $\rho_\alpha^0 = \lambda \rho_2 $ or $\rho_\beta^0 = (1-\lambda)\rho_2$
\item[•] The solution to 
\begin{equation}
\begin{cases}
{\Gamma_\alpha\rho_\alpha c_{v\alpha}} = {\Gamma_\beta\rho_\beta c_{v\beta}} \\
\dfrac{1}{\rho_2} = \dfrac{\lambda}{\rho_\alpha} + \dfrac{1-\lambda}{\rho_\beta} \;, 
\end{cases}
\end{equation}
as suggested by Kapila et al. \cite{Kapila_2007}.
\end{itemize}

\subsubsection{Two-dimensional root-finding algorithm}

The application of temperature and pressure equilibrium between the reactants and the products of each reactive mixture yields the following non-linear system of equations:
\begin{equation}\label{eqn:Root2d_Teq}
T_{\alpha_i}(p,\rho_{\alpha_i}) = T_{\beta_i}(p,\rho_{\beta_i}), \hspace{0.1cm} \rm{for} \ i=1,2 \;,\\
\end{equation}
which must be solved along with the mixture rules \eqref{eqn:MixRulesDensityExt}; the temperatures are calculated using \eqref{eqn:T_off_curve}. Solving \eqref{eqn:Root2d_Teq} corresponds to finding the root of the function ${\bf{F}}_{T_{eq}}$, defined as
\begin{equation}
{\bf{F}}_{T_{eq}} = \begin{pmatrix} F^{(1)}_{T_{eq}} \\ F^{(2)}_{T_{eq}} \end{pmatrix} = \begin{pmatrix} T_{\alpha_1}(p,\rho_{\alpha_1}) - T_{\beta_1}(p,\rho_{\beta_1}) \\ T_{\alpha_2}(p,\rho_{\alpha_2}) - T_{\beta_2}(p,\rho_{\beta_2}) \end{pmatrix} \;.
\end{equation}
If the task is to invert the mixed equation of state \eqref{eqn:MixedEOS_pressure} and compute the internal energy when the pressure is known, equations \eqref{eqn:Root2d_Teq} are decoupled and can be solved separately applying to each equilibrium equation the previously described one-dimensional root-finding technique. Equation \eqref{eqn:MixedEOS_energy} can be then used to compute the energy once the values of the two independent variables have been found. If, instead, the pressure is unknown (but $\rho e$ is known), expression \eqref{eqn:MixedEOS_pressure} can be used to eliminate $p$ from both equations. In this case, however, the dependence of $p$ on the densities of all constituents renders the two equations fully coupled. 

As in the one-dimensional case, when solving ${\bf{F}}_{T_{eq}} = {\bf{0}}$ numerically, consideration must first be given to the choice of the independent variables. These are selected according to the same criterion introduced in Sect.~\ref{Sect:RootFinding1D} and based on the values of the mass fraction variables $\lambda_1$ and $\lambda_2$. The main problems associated with the numerical solution of this two-dimensional root-finding problem are: (i)
the number of solutions in not known a priori; (ii) the limited applicability of the EoS models, leading to mathematical solutions  which may be physically invalid (resulting for instance in a negative sound speed); (iii) the absence of a robust multi-dimensional root-finding technique that can be applied to virtually any problem \cite{Press_2007}. To address all these issues, a numerical strategy has been developed in this work which is tailored specifically for the problem under consideration and takes advantage of physical and thermodynamic understanding. The proposed root-finding technique is based on the combined use of the standard Newton-Raphson method and a direct search optimisation method which aims to ensure the robustness of the overall algorithm. The resulting method can be summarised as follows:
\begin{itemize}[leftmargin=0.5cm]
\item[]{\bf{Step 1: Initial guess}}. As in the one-dimensional root-finding technique described above, the availability of a good starting value for each independent variable is fundamental to guarantee the fast convergence of the method to a physical solution. Also in this case, the best guess value is represented by the converged solution at the previous time step.
\item[]{\bf{Step 2: Search domain}}. Again, as in the one-dimensional technique, it is very important to define a search domain by determining a lower and upper limit for each independent variable. As before, the lower limits are deduced from the requirement that the density of each constituent must be positive. As for the determination of the upper limits, we have adopted the strategy proposed by Wilkinson et al. \cite{Wilkinson2017}, where the maximum density is obtained as the one for which the temperature equals zero at a pressure of $1.5$ times the Von Neumann pressure \cite{Lee_book}. As observed in Wilkinson et al.\cite{Wilkinson2017}, this somewhat arbitrary choice was found to eliminate any unphysical behaviour in the large density limit, while at the same time ensuring that the physical solution was never accidentally excluded from the search domain. 
\item[]{\bf{Step 3: Problem simplification}}. For most points on the computational grid the root-finding problem can be considerably simplified:
\begin{itemize}
\item At grid points in the domain away from interfaces (i.e.\ $z_1$ is within a tolerance of $0$ or $1$) and reaction zones (i.e.\ both $\lambda_1$ and $\lambda_2$ are within a tolerance of $0$ or $1$), only one component is present and therefore no root-finding is required as the mixture EoS \eqref{eqn:MixedEOS_energy} can be evaluated explicitly. 
\item A similar situation occurs at grid points far away from reaction zones, but near material interfaces. Here, each explosive phase is composed of a single constituent and, as a result, the density of the existing component is equal to the overall density  of the corresponding explosive phase.
\item Another special case occurs for grid points in the proximity of reaction zones, but away from material interfaces. In this case,  
only one phase is present and the one-dimensional root-finding technique described above is sufficient to retrieve the densities of the constituents of the existing explosive phase.
\end{itemize}
\item[]{\bf{Step 4: Iterative algorithm}}. 
At grid points in the proximity of both material interfaces and reaction zones, the full two-dimensional technique is applied. At every iteration, the function ${\bf{F}}_{T_{eq}}$ is evaluated, along with the associated Jacobian matrix $\underline{\underline{\bf{J_F}}}{(\bf{x})}= \frac{\partial {{\bf{F}}_{T_{eq}}}}{\partial{\bf{x}}}$, and an iteration of the Newton-Raphson method is carried out:
\begin{equation}
{{\bf{x}}^{(k+1)}} = {{\bf{x}}^{(k)}} - \underline{\underline{\bf{J_F}}}^{-1}{({{\bf{x}}^{(k)}})}{\bf{F}}_{T_{eq}}{({{\bf{x}}^{(k)}})} \;.
\end{equation}
In order to be accepted, the Newton-Raphson iteration has to satisfy two important requirements:
\begin{itemize}
\item[i)] the new value ${{\bf{x}}^{(k+1)}}$ must fall within the search space;
\item[ii)] the Jacobian matrix $\underline{\underline{\bf{J_F}}}{(\bf{x})}$ evaluated at ${{\bf{x}}^{(k)}}$ must not be ill-conditioned. To detect ill-conditioning we have introduced the classic condition number
\begin{equation}
Cond(\underline{\underline{\bf{J_F}}}{(\bf{x})}) = \norm{\underline{\underline{\bf{J_F}}}}_2\norm{\underline{\underline{\bf{J_F}}}^{-1}}_2 \;,
\end{equation}
with 
\begin{equation}
\norm{\underline{\underline{\bf{J_F}}}}_2 = \max_{\bf{x} \in \mathbb{R}^2, \bf{x} \neq 0} \dfrac{\norm{\underline{\underline{\bf{J_F}}}\bf{x}}_2}{\norm{\bf{x}}_2} \;,
\end{equation}
and specified a condition number threshold $\delta_{cond}$ (in this work we took $\delta_{cond} = \SI{1e8}{}$).
\end{itemize}
When the Newton-Raphson iteration is rejected (on the basis of the above requirements), some iterations (typically $10-20$) are performed with a direct search optimisation method minimising the following objective function 
\begin{equation}
f^{obj}{(\bf{x})} = \sqrt{\sum \limits_i F^{(i)}_{T_{eq}}{(\bf{x})}} \;.
\end{equation}
In particular, the Nelder Mead method \cite{NelderMead} has been chosen in this work due to its inherent robustness and simplicity of implementation. The algorithm terminates when both the following stopping criteria are satisfied: 
\begin{equation}
\begin{cases}
f_{MIN}({{\bf{x}}^{(k)}}) < \epsilon_f \;, \\
\norm{{{\bf{x}}^{(k)}} - {{\bf{x}}^{(k-1)}}}_2 < \epsilon_x \;,
\end{cases}
\end{equation}
where $\epsilon_f$ and $\epsilon_x$ are tolerance values provided by the user. In this work we set $ \epsilon_f = \SI{1e-6}{}$ and $ \epsilon_x = \SI{1e-9}{}$. When the algorithm does not converge or converges to an unphysical solution, a different starting value ${\bf{x}}^0=[x_1^0, x_2^0]$ is tried, as in the one-dimensional algorithm presented above.
\end{itemize}

\section{Numerical Results}\label{Sect:Num_results}

In this section, we present a selection of numerical results to illustrate the full capability of our extended model and demonstrate its applicability to scenarios involving two reactive materials in direct contact with each other. For all test problems, the numerical solutions are obtained using the proposed formulation with the MUSCL-Hancock finite volume method and the HLLC approximate Riemann solver. Tables~\ref{Tab:EOS} and \ref{Tab:RateLaw} display the equation of state and reaction rate parameters for all materials considered in the test problems of this section. Unless otherwise stated, a CFL coefficient $C_{cfl}=0.8$ is used for all computations. For thoroughness and code validation purposes, some simulations of two-component and three-component benchmark test problems have also been performed and are briefly presented in the next section.

%
%
%
\begin{table}[t]
  \centering
	\renewcommand{\arraystretch}{1.25} 	
	\resizebox{0.75\textwidth}{!}{%
	\begin{tabular}{ l l | c c c c c c c c } 
    \hline
    {\large{\bf{EOS}}} & {\large{Material}} & & & & & & & & \Tstrut\Bstrut\\ \hline 
    {\bf{Cochran-Chan}} &  & $\Gamma$ & $\mathcal{A}$ & $\mathcal{B}$ & $\mathcal{E}_1$ & $\mathcal{E}_2$ & $\rho_0$ & $c_v$ & $Q$\Tstrut\Bstrut\\ \hline
    &  & $[-]$ & $[\SI{}{\pascal}]$ & $[\SI{}{\pascal}]$ & $[-]$ & $[-]$ & $[\SI{}{\kilo\gram\meter^{-3}}]$ & $[\SI{}{\meter^2\second^{-2}\kelvin^{-1}}]$ & $[\SI{}{\meter^2\second^{-2}}]$ \Tstrut\Bstrut\\ \hline
    & Copper & 2.0 & \SI{145.67e9}{} & \SI{147.75e9}{} & 2.99 & 1.99 & 8900 &393 & - \\ \hline
     & TNT Prod. & 0.25 & \SI{8.545e11}{} & \SI{2.05e10}{} & 4.6 & 1.35 & 1840 &815 & - \\ \hline
     & Nitromethane & 1.19 & \SI{0.8192e9}{} & \SI{1.508e9}{} & 4.53 & 1.421 & 1134 &1714 & \SI{4.485e6}{} \\ \hline
     {\bf{JWL}} &  & $\Gamma$ & $\mathcal{A}$ & $\mathcal{B}$ & $\mathcal{R}_1$ & $\mathcal{R}_2$ & $\rho_0$ & $c_v$ & $Q$\\ \hline
     &  & $[-]$ & $[\SI{}{\pascal}]$ & $[\SI{}{\pascal}]$ & $[-]$ & $[-]$ & $[\SI{}{\kilo\gram\meter^{-3}}]$ & $[\SI{}{\meter^2\second^{-2}\kelvin^{-1}}]$ & $[\SI{}{\meter^2\second^{-2}}]$ \Tstrut\Bstrut\\ \hline
     & LX-17 React. & 0.8938 & $\SI{778.1e11}{}$ & $\SI{-0.5031e10}{}$ & 11.3 & 1.13 & 1905 &1305.5 & - \Tstrut\Bstrut\\ \hline
     & LX-17 Prod. & 0.5 & $\SI{14.8105e11}{}$ & $\SI{6.379e10}{} $& 6.2 & 2.2 & 1905 &524.9 & $\SI{3.620e6}{}$ \Tstrut\Bstrut\\ \hline
     & PETN React. & 0.5675 & $\SI{202.8e11}{}$ & $\SI{-0.3752e10}{}$ & 10.0 & 1.0 & 1778 &1528.4 & - \Tstrut\Bstrut\\ \hline
     & PETN Prod. & 0.5 & $\SI{10.3216e11}{}$ & $\SI{9.057e10}{} $& 6.0 & 2.6 & 1778 &562.4 & $\SI{6.074e6}{}$ \Tstrut\Bstrut\\ \hline
 & UF-TATB React. & 0.8938 & $\SI{63.207e11}{}$ & $\SI{-0.4472e10}{}$ & 11.3 & 1.13 & 1800 &1381.7 & - \Tstrut\Bstrut\\ \hline
     & UF-TATB Prod. & 0.5 & $\SI{12.05026e11}{}$ & $\SI{6.02513e10}{} $& 6.2 & 2.2 & 1800 &555.6 & $\SI{3.690e6}{}$ \Tstrut\Bstrut\\ \hline
\end{tabular} }
\vspace{2mm}
\caption{Equation of state  parameters for the materials considered in the test problems of this chapter. The nitromethane is modelled using the Cochran-Chan equation of state, as in Michael and Nikiforakis \cite{Michael_Nikiforakis:2019_a, Michael_Nikiforakis:2019_b} and Mi et al. \cite{MIMINI19}. The JWL parameters for LX-17, PETN (single crystal) and UTATB are taken from Tarver et al. \cite{Tarver2005, Tarver_Breithaupt}.}
\label{Tab:EOS}
\end{table}
\begin{table}[t]
  \centering
	\renewcommand{\arraystretch}{1.25} 	
	\resizebox{0.6\textwidth}{!}{%
	\begin{tabular}{ | c | l c | l c | l c |} 
    \hline
    {{\bf{Material}}} & \multicolumn{6}{c|}{\bf{Reaction Rate Law}} \\ \hline 
    \multirow{2}{*}{Nitromethane} & \multicolumn{6}{c|}{{Single-step Arrhenius}} \\ \cline{2-7}
& \multicolumn{1}{c}{$T_A [\SI{}{\kelvin}]$} & \multicolumn{1}{c|}{11350} & \multicolumn{1}{c}{$A \ [\SI{}{\second^{-1}}]$} &\multicolumn{3}{l|}{ \SI{2.6e9}{}} \\ \hline
    \multirow{7}{*}{LX-17} & \multicolumn{6}{c|}{{$I\&G$}} \\ \cline{2-7}  
                           & \multicolumn{2}{c|}{$\mathcal{R}_{IG}$} & \multicolumn{2}{c|}{$\mathcal{R}_{G1}$}& \multicolumn{2}{c|}{$\mathcal{R}_{G2}$}\\ \cline{2-7} 
                           & $I \ [\SI{}{\second^{-1}}]$ & \SI{4.0e12}{}& $\tilde{G_1} \ [\SI{}{\giga\pascal^{-3}\second^{-1}}]$ & 4500 & $\tilde{G_2}  \ [\SI{}{\giga\pascal^{-1}\second^{-1}}]$ & \SI{3.0e5}{} \\
                           & $a \ [-]$ & 0.22 & $b \ [-]$ & 0.667 & $c  \ [-]$ & 0.667\\
                           & $d \ [-]$ & 1 & $e \ [-]$ & 0.667 & $g  \ [-]$ & 0.667\\
                           & $x \ [-]$ & 7 & $y \ [-]$ & 3 & $z  \ [-]$ & 1\\
                           & $\Phi_{ig,max} \ [-]$ & 0.02 & $\Phi_{G_1,max} \ [-]$ & 0.8 & $\Phi_{G_2,max}  \ [-]$ & 0.8 \\ \hline
\multirow{7}{*}{PETN} & \multicolumn{6}{c|}{{$I\&G$}} \\ \cline{2-7}  
                           & \multicolumn{2}{c|}{$\mathcal{R}_{IG}$} & \multicolumn{2}{c|}{$\mathcal{R}_{G1}$}& \multicolumn{2}{c|}{$\mathcal{R}_{G2}$}\\ \cline{2-7} 
                           & $I \ [\SI{}{\second^{-1}}]$ & \SI{4.0e8}{}& $\tilde{G_1} \ [\SI{}{\giga\pascal^{-1}\second^{-1}}]$ & 1000 & $\tilde{G_2}  \ [\SI{}{\giga\pascal^{-2}\second^{-1}}]$ & \SI{1.5e5}{} \\
                           & $a \ [-]$ & 0 & $b \ [-]$ & 0.667 & $c  \ [-]$ & 0.667\\
                           & $d \ [-]$ & 0.01 & $e \ [-]$ & 0.667 & $g  \ [-]$ & 0.667\\
                           & $x \ [-]$ & 9 & $y \ [-]$ & 1 & $z  \ [-]$ & 2\\
                           & $\Phi_{ig,max} \ [-]$ & 0.01 & $\Phi_{G_1,max} \ [-]$ & 0.01 & $\Phi_{G_2,max}  \ [-]$ & 0.01 \\ \hline
\multirow{7}{*}{UF-TATB} & \multicolumn{6}{c|}{{$I\&G$}} \\ \cline{2-7}  
                           & \multicolumn{2}{c|}{$\mathcal{R}_{IG}$} & \multicolumn{2}{c|}{$\mathcal{R}_{G1}$}& \multicolumn{2}{c|}{$\mathcal{R}_{G2}$}\\ \cline{2-7} 
                           & $I \ [\SI{}{\second^{-1}}]$ & \SI{4.0e8}{}& $\tilde{G_1} \ [\SI{}{\giga\pascal^{-2}\second^{-1}}]$ & \SI{2.2e5}{} & $\tilde{G_2}  \ [\SI{}{\giga\pascal^{-1}\second^{-1}}]$ & \SI{6.0e5}{} \\
                           & $a \ [-]$ & 0.214 & $b \ [-]$ & 0.667 & $c  \ [-]$ & 0.667\\
                           & $d \ [-]$ & 1 & $e \ [-]$ & 0.667 & $g  \ [-]$ & 0.667\\
                           & $x \ [-]$ & 7 & $y \ [-]$ & 2 & $z  \ [-]$ & 1\\
                           & $\Phi_{ig,max} \ [-]$ & 0.071 & $\Phi_{G_1,max} \ [-]$ & 1.0 & $\Phi_{G_2,max}  \ [-]$ & 0.8 \\ \hline
\end{tabular}}
\vspace{2mm}
\caption{Reaction rate law parameters for the materials considered in the test problems of this chapter. The $I\&G$ parameters for LX-17, PETN (single crystal) and UTATB are taken from Tarver et al. \cite{Tarver2005, Tarver_Breithaupt}. The reaction process in nitromethane is modelled by a single-step temperature-dependent Arrhenius rate law, as in Michael and Nikiforakis \cite{Michael_Nikiforakis:2019_a, Michael_Nikiforakis:2019_b} and Mi et al. \cite{MIMINI19}}.
\label{Tab:RateLaw}
\end{table}
\clearpage
\subsection{Code validation}\label{Sect:codevalidation}

To validate our implementation, we have performed simulations of two-component and three-component benchmark tests and compared our numerical solutions with those of published work or, where available, with exact solutions. This is done for thoroughness and to verify that, for problems consisting of no more than three components, our extended formulation provides virtually identical results to those obtained using the MiNi16 model \cite{MINI16}. To simulate these scenarios, some materials appearing in the full system proposed in Sect.~\ref{Sect:Extend Formulation} are selectively rendered inactive, following the strategy outlined in Sect.~6 of Michael and Nikiforakis \cite{MINI16}; in this way one can simulate, for example, (i) an explosive interacting with an inert material, (ii) a two-fluid mixture, (iii) two materials separated by a sharp interface, or (iv) a simple single-component flow.

\subsubsection{Copper-detonation products shock tube}

\begin{figure*}
  \centering
  \subfloat{\includegraphics[width=.33\linewidth]{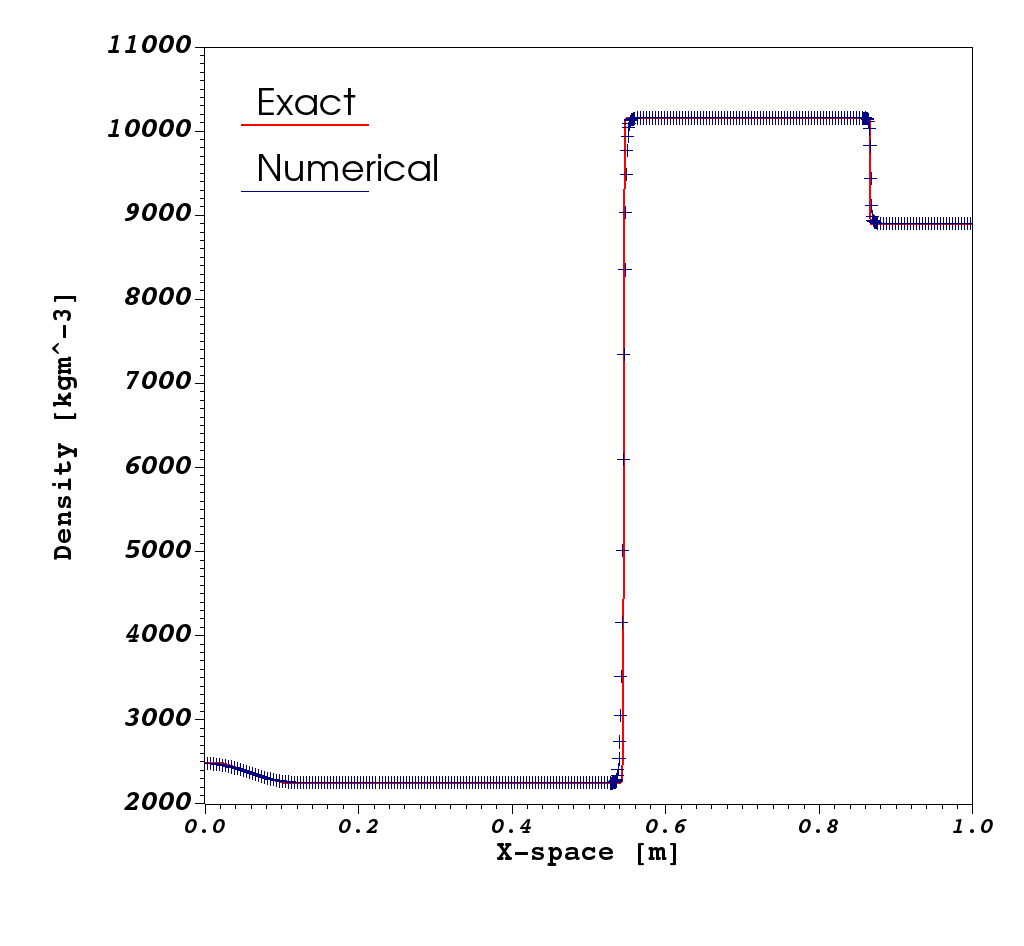}}\quad
\subfloat{\includegraphics[width=.33\linewidth]{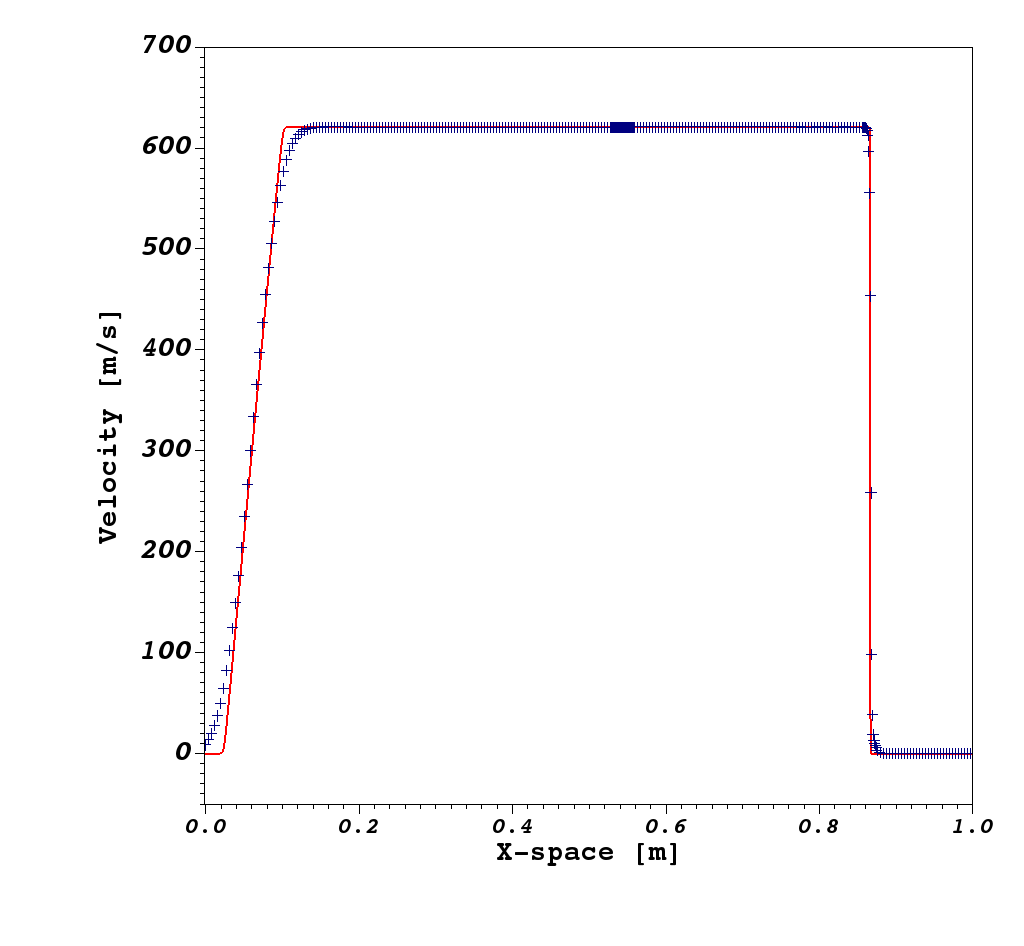}} \\[-0.46cm]
 \subfloat{\includegraphics[width=.33\linewidth]{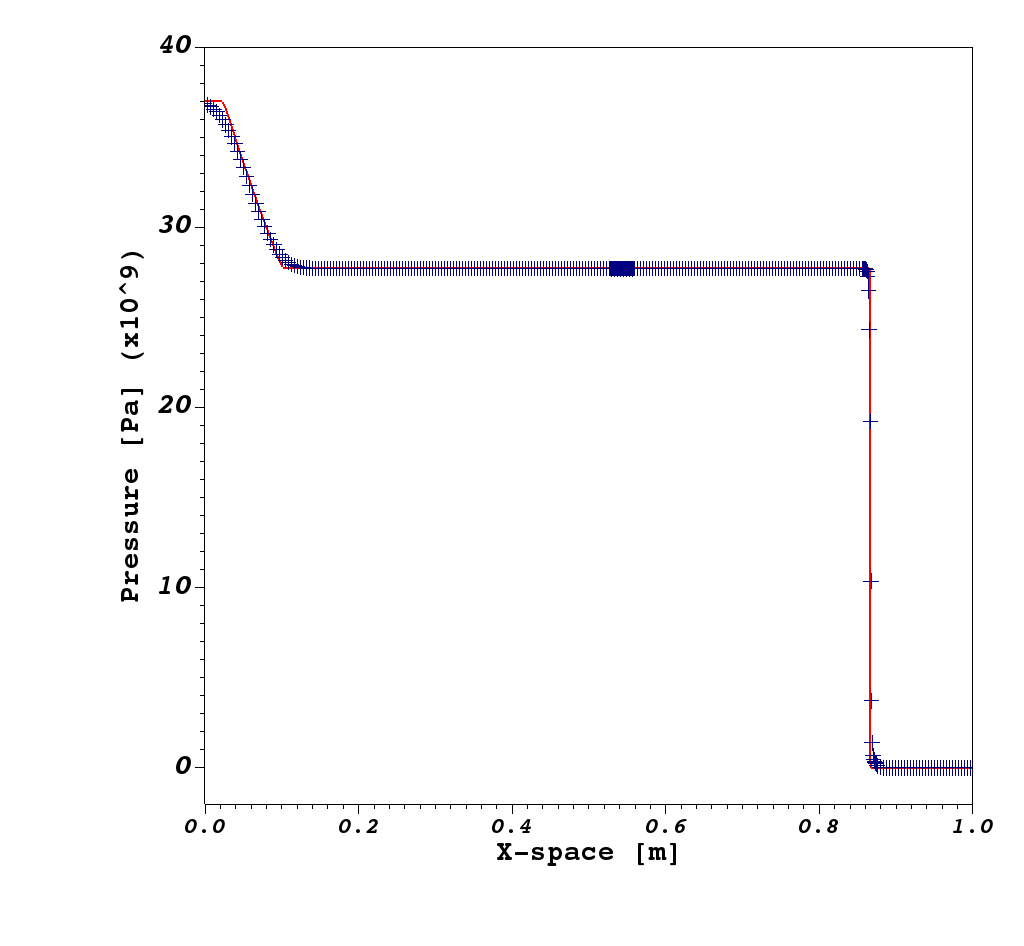}}\quad
\subfloat{\includegraphics[width=.33\linewidth]{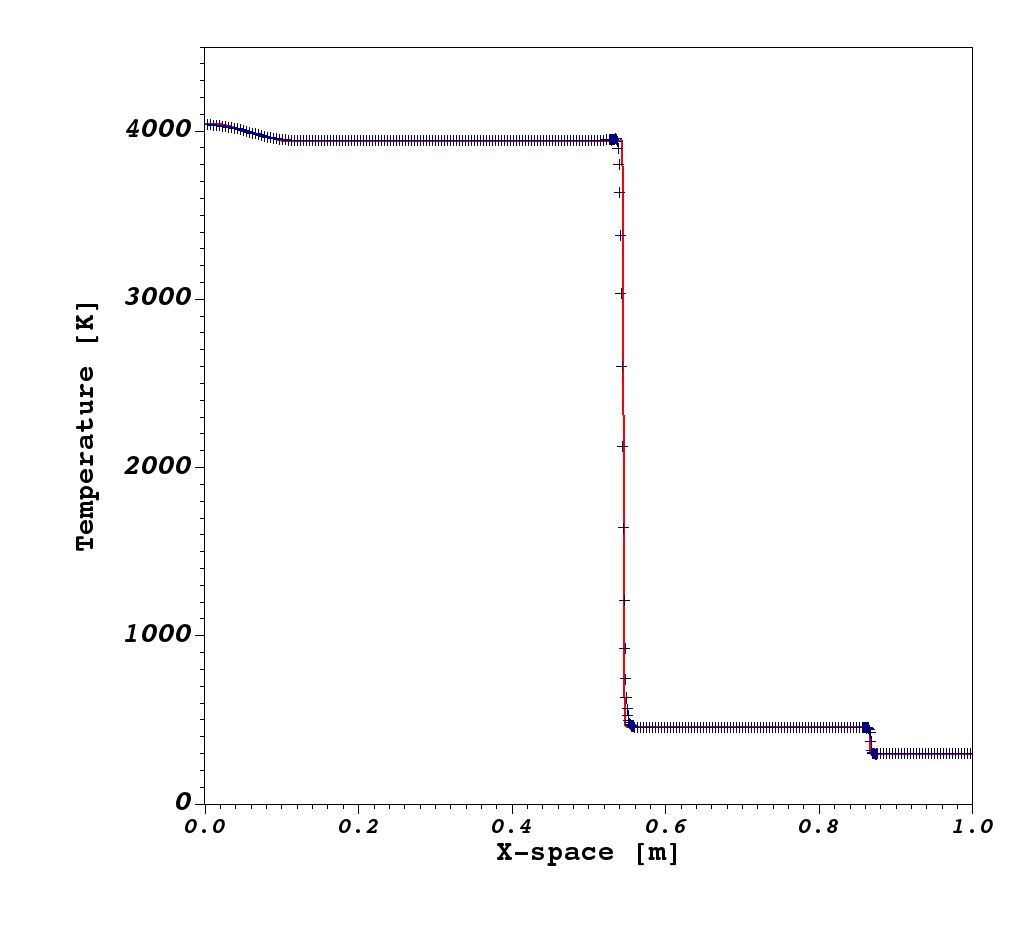}}
  \caption{\label{Fig:Copper1D} Results for the copper-detonation products shock tube problem. Profiles for density, velocity, pressure and temperature are shown at time $t=\SI{73}{\micro\second}$. The red line is the exact solution and the blue crosses represent the numerical solution.}
\end{figure*}

This is a two-component inert shock-tube test featuring a high-pressure gas region (TNT detonation products) in contact with a copper plate \cite{Saurel:1999a}. Solution profiles for density, velocity, pressure and temperature across the complete wave structure are illustrated in Fig.~\ref{Fig:Copper1D} at time $t=\SI{73}{\micro\second}$. The exact solution consists of a rarefaction wave propagating to the left in the TNT products and a strong shock wave travelling to the right in the copper, along with a material interface in between. The numerical solutions are obtained using the same setup and EoS parameters as in Saurel and Abgrall \cite{Saurel:1999a}. A base coarse grid of ${\rm N}=256$ computing cells is used and a single level of refinement, with a refinement factor of four, is employed to refine the shock and the material interface. By comparing the computed solutions with the exact Riemann solution, good agreement is observed. Also, an excellent agreement is observed between the current simulation result and that reported by Saurel and Abgrall \cite{Saurel:1999a} and Michael and Nikiforakis \cite{MINI16}.

%
%

\subsubsection{Void collapse in liquid reactive nitromethane}

\begin{figure}[tb!]
  \centering
    \includegraphics[width=0.475\textwidth]{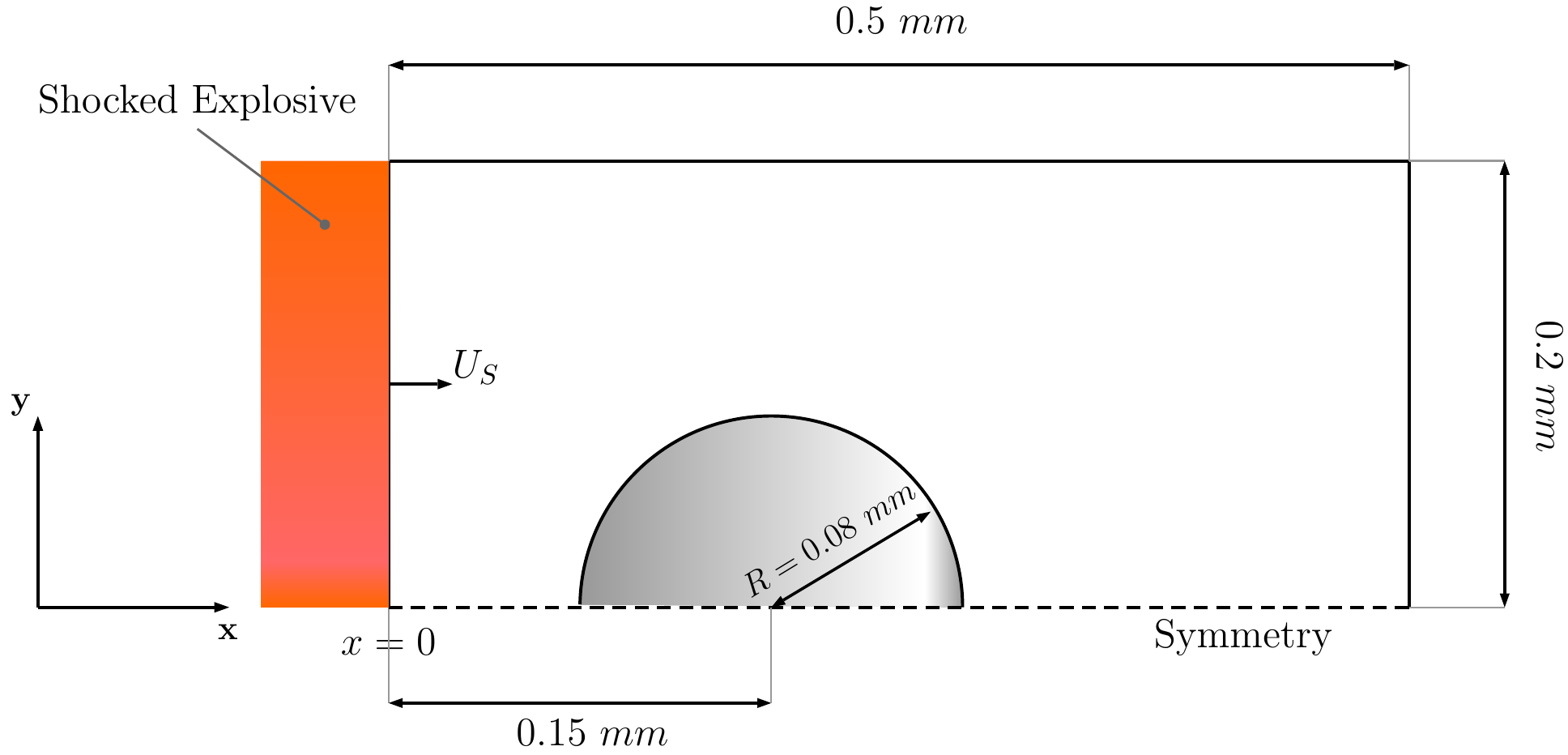} 
  \caption{Schematic illustration of the computational setup for the shock-induced collapse of an isolated air cavity (gray shaded area) in liquid nitromethane.}
\label{Fig:NitroBubble_schematic}
\end{figure}

\begin{figure}[tb!]
  \centering
  \includegraphics[width=0.33\textwidth, angle =270]{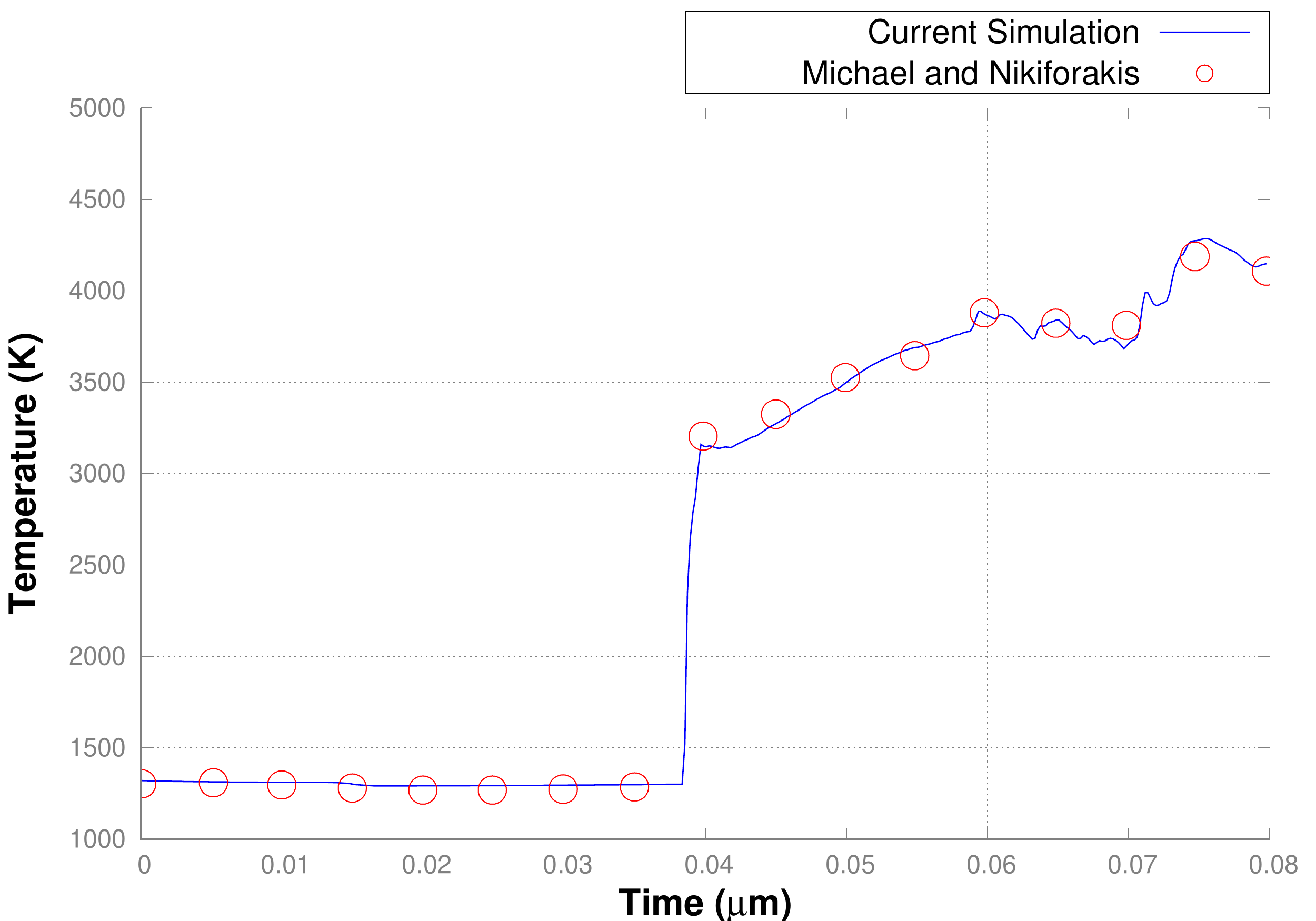}
  \caption{Time evolution of the maximum hot-spot temperature in liquid nitromethane. The numerical solution obtained in this work (blue curve) is compared to that reported in Michael and Nikiforakis \cite{Michael_Nikiforakis:2019_b} (red circles).}
\label{Fig:MaxTempNitroCollapse}
\end{figure}

As a second validation test, we have simulated an isolated air cavity collapsing in liquid reactive nitromethane. Fig.~\ref{Fig:NitroBubble_schematic} schematically shows the initial configuration of the computational domain for this test. The dimensions of the initial features and the simulation parameters (including EoS and reaction rate models) are the same as those used by Michael and Nikiforakis \cite{Michael_Nikiforakis:2019_b}. An air cavity with an initial diameter of $\SI{160}{\micro\meter}$ is subjected to a $\SI{10.98}{\giga\pascal}$ incident shock wave. The calculation was performed at a grid resolution of $\Delta x = \Delta y = \SI{0.3125}{\micro\meter}$, consistent with that used by Micheal and Nikiforakis \cite{Michael_Nikiforakis:2019_b}. In Fig.~\ref{Fig:MaxTempNitroCollapse} the time evolution of the maximum nitromethane temperature across the entire domain is illustrated and directly compared to the numerical solution reported in Fig.~14 of Michael and Nikiforakis \cite{Michael_Nikiforakis:2019_b}. Excellent agreement is observed between the results, meaning that the extension performed in this work has not impaired the key features of the original MiNi16 model.

\subsection{Numerical results in the presence of two reactive materials}

\subsubsection{Detonation diffraction around a rigid corner}

\begin{figure}
  \centering
    \includegraphics[height=0.295\textwidth]{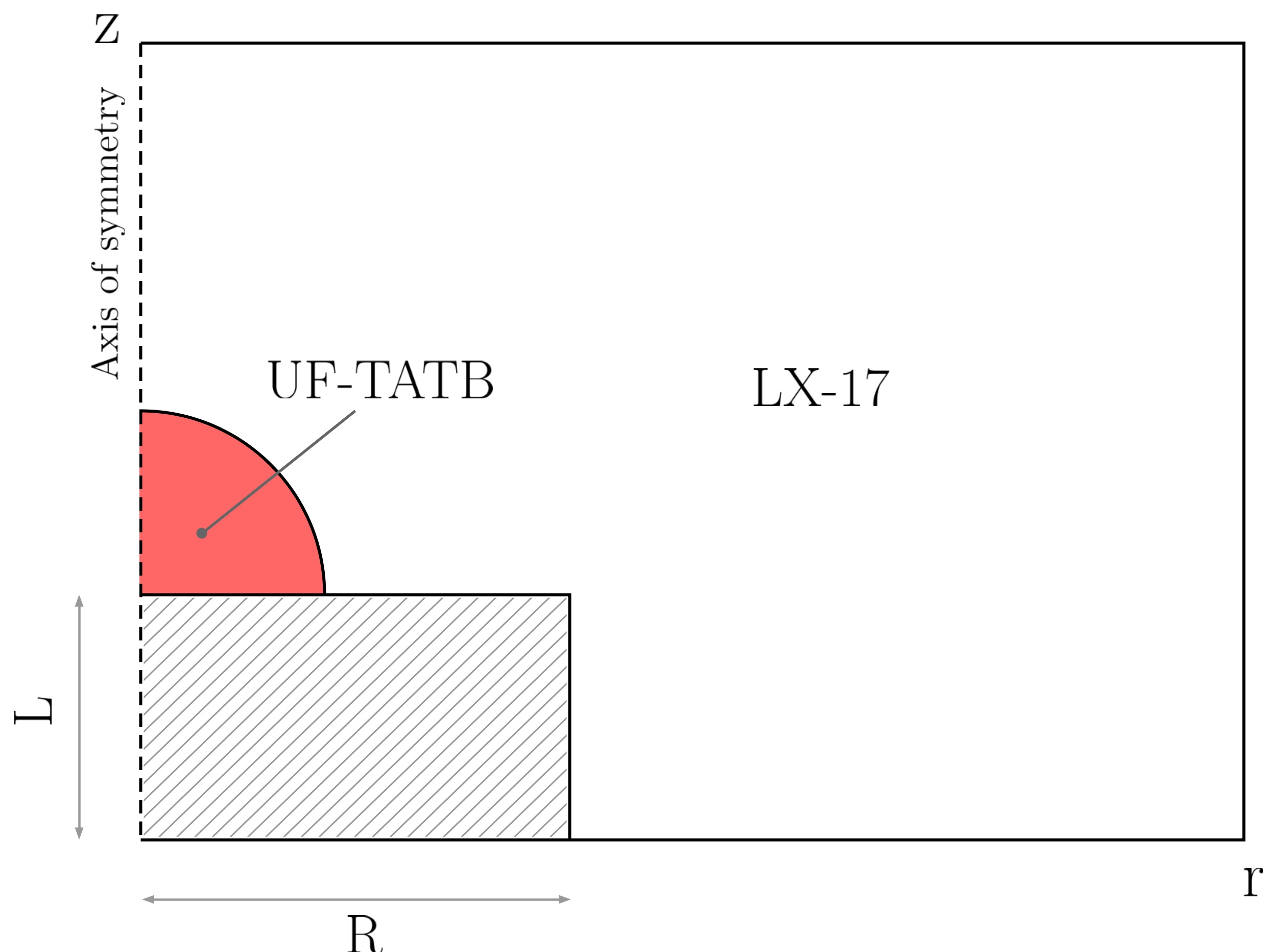} 
  \caption{Two-dimensional illustration of the hockey puck configuration.  It consists of a cylindrical LX-17 sample with a coaxial cylindrical cavity of radius $R=\SI{19.05}{\milli\meter}$ and length $L=\SI{15}{\milli\meter}$. The detonation is initiated by a hemispherical region of radius $\SI{7.68}{\milli\meter}$ filled with UF-TATB products at the pressure of $\SI{31.46}{\giga\pascal}$ .}
\label{Fig:Diffraction_Scheme}
\end{figure}

As a first test involving two distinct explosive materials, we consider a hockey puck configuration that follows the experiment by Souers et al. \cite{Souers} and consists of a cylindrical LX-17 sample from which a shorter, coaxial cylinder has been removed to form a well. A schematic of the hockey puck geometry examined here is shown in Fig.~\ref{Fig:Diffraction_Scheme}. The setup is axisymmetric and the LX-17 charge is initiated by a hemispherical region of radius $\SI{7.68}{\milli\meter}$, embedded within the LX-17 at the base of the cavity and filled with high pressure Ultrafine TATB (UF-TATB) products. Note that this is different from the original study by DeOliveira \cite{DeOliveira}, where LX-17 products at a very high pressure are used to initiate the main sample of explosive. The complete initial conditions for this test problem are shown in Table \ref{Tab:IC_HockeyPuck}.
\begin{figure}
  \centering
  \renewcommand{\thesubfigure}{a}
  \subfloat[]{\includegraphics[width=.375\linewidth]{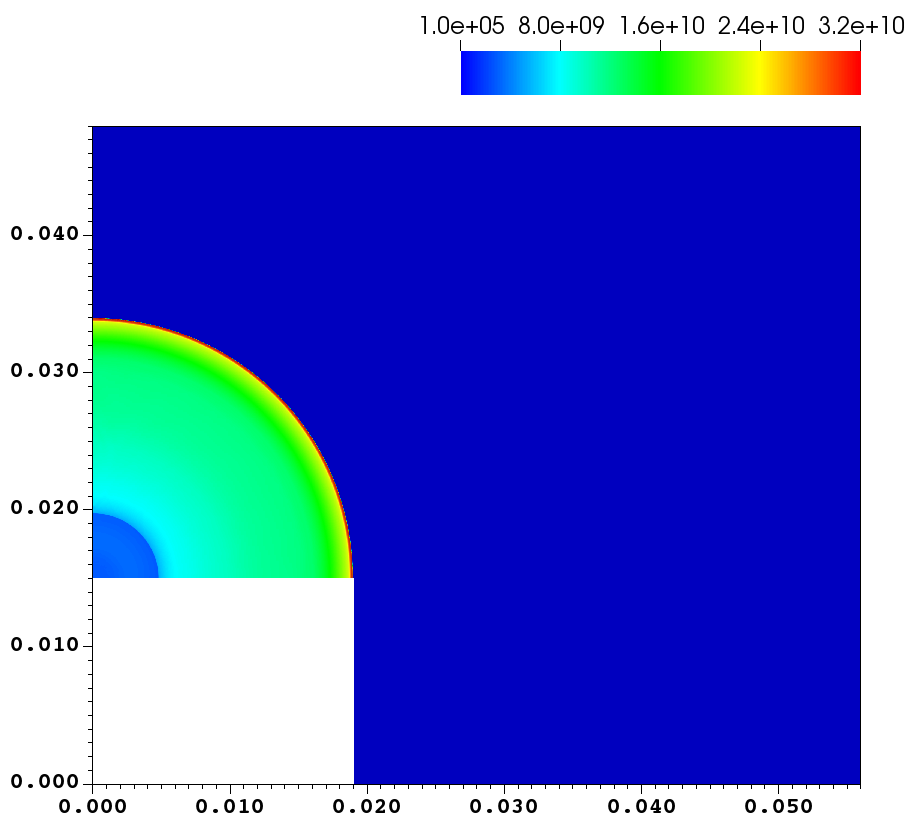}}\quad
   \renewcommand{\thesubfigure}{b}
\subfloat[]{\includegraphics[width=.375\linewidth]{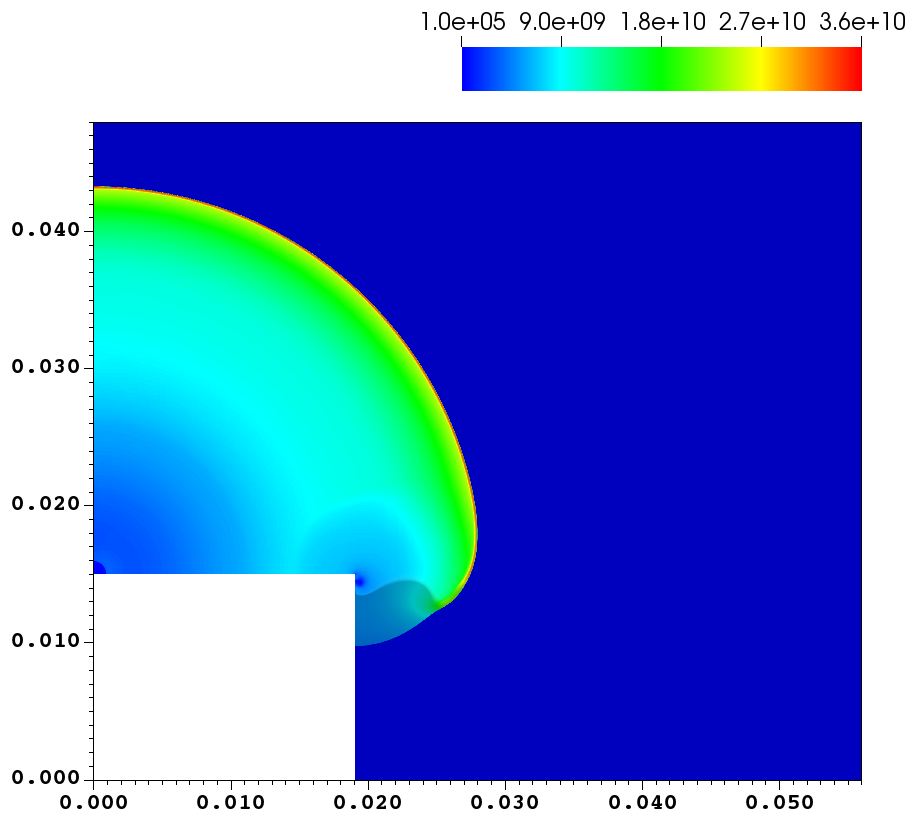}} \\
 \renewcommand{\thesubfigure}{c}
 \subfloat[]{\includegraphics[width=.375\linewidth]{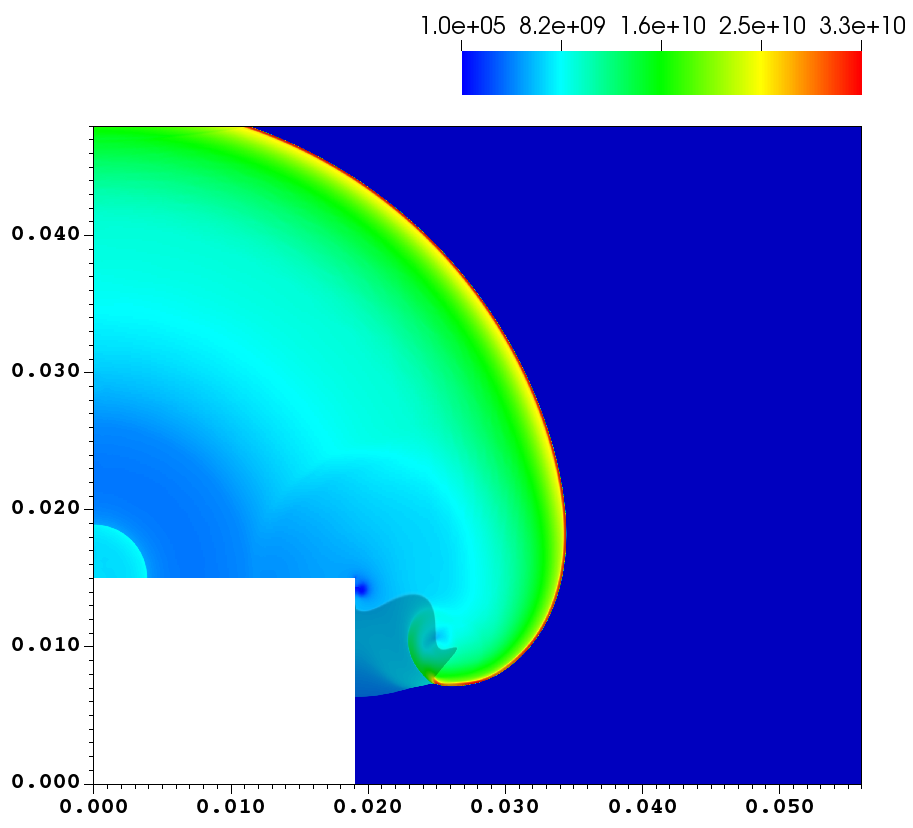}}\quad
  \renewcommand{\thesubfigure}{d}
\subfloat[]{\includegraphics[width=.375\linewidth]{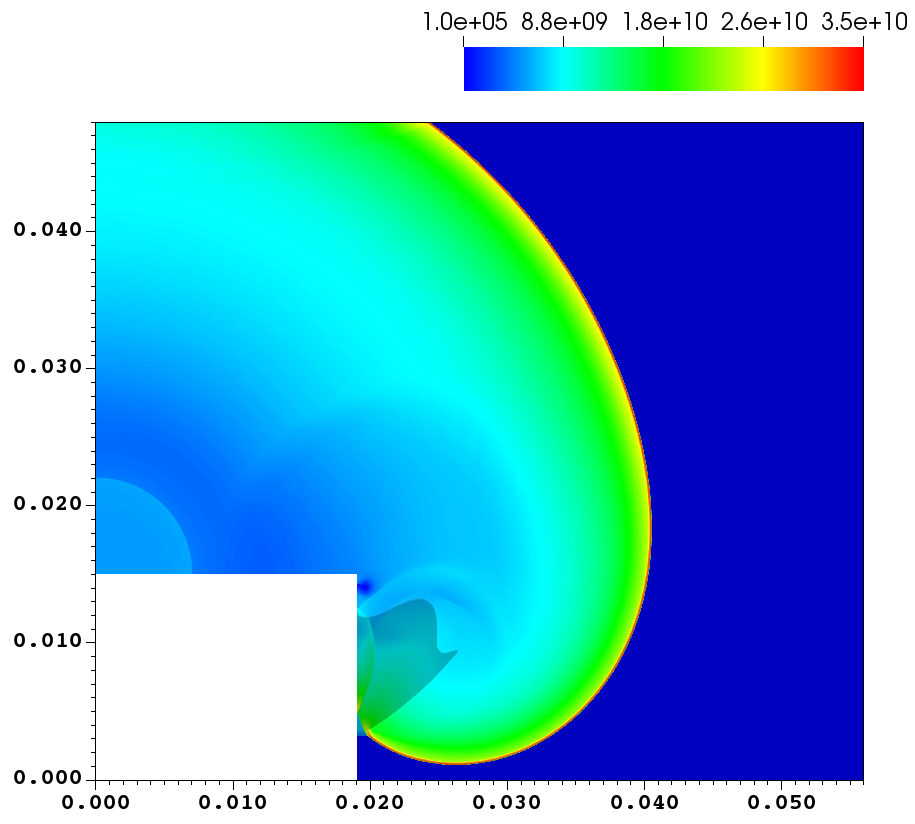}}
\vspace*{-0.2cm}
  \caption{\label{Fig:HockeyPuck} Numerical solution of pressure field for the hockey puck configuration at times (a) $t=\SI{1.6}{\micro\second}$, (b) $t=\SI{2.9}{\micro\second}$, (c) $t=\SI{3.8}{\micro\second}$ and (d) $t=\SI{4.6}{\micro\second}$. The shaded region indicates areas where the explosive has not reacted (dead zone). }
\end{figure}

The rapid expansion of the UF-TATB products leads to the prompt formation of a hemispherical detonation that propagates into the main explosive charge and subsequently diffracts around the right angled corner. In the numerical setup of this problem, the boundary of the well is treated as rigid, as in the work of DeOliveira et al. \cite{DeOliveira}. A reflective boundary condition is prescribed along the axis of symmetry $y=0$, while the remaining domain boundaries are taken to be transmissive. The equations of state (JWL) and reaction rate law (Ignition and Growth) parameters for both explosive materials (LX-17 and UF-TATB) are given in Tables \ref{Tab:EOS} and \ref{Tab:RateLaw}. In order to capture desensitisation effects and the formation of dead zones, the desensitisation model proposed by DeOliveira et al. \cite{DeOliveira} is included in the formulation with the same parameters used therein. The computational domain is discretised with a coarse mesh of size $\SI{160}{\milli\meter} $ and two levels of refinement, with refinement factors $\times 4 $ and $\times 2 $ are used to adaptively increase the resolution around the detonation front, material interfaces and rigid boundaries; this results in an effective resolution of $\Delta x = \Delta y = \SI{20}{\micro \meter}$ which is consistent with that used by Ioannou and Nikiforakis \cite{IoannouII:2020}.

Numerical results are shown in Fig.~\ref{Fig:HockeyPuck}. As the detonation expands around the right angled corner at $t \sim \SI{1,6}{\micro\second}$ (see frame (a) of Fig.~\ref{Fig:HockeyPuck}), it undergoes diffraction and weakens, causing the reactions to locally reduce, consequently leading to the formation of a curved inert shock that compresses and desensitises the explosive (see frame (b) of Fig.~\ref{Fig:HockeyPuck}). In the region away from the corner the detonation front continues to propagate undisturbed as a spherical wave. As the detonation travels faster than the inert shock, it eventually reaches the vertical wall of the rigid well, producing a region within the explosive sample where no reactions have occurred; this results in a dead zone that  is observed to persist for the rest of the simulation. The results are in good agreement with the study by DeOliveira et al. \cite{DeOliveira} and the numerical solutions reported in Ioannou and Nikiforakis \cite{IoannouII:2020}, where high pressure LX-17 products are used to initiate the main sample of explosive. Indeed, the use of UF-TATB products does not qualitatively affect the overall results, but illustrates the applicability of the proposed model to scenarios involving two distinct explosive materials. This test also demonstrates the robustness of the proposed model across the material interface between reacting materials, which is seen to be oscillation-free.
\begin{table}
  \vspace{1.5mm}
  \centering
	\renewcommand{\arraystretch}{1.25} 	
	\resizebox{0.485\textwidth}{!}{%
	\begin{tabular}{ c | c c c c c c c c} 
    \hline
    \multirow{2}{*}{Material} & $ \rho_{\alpha_1} = \rho_{\beta_1}$ & $\rho_{\alpha_2} = \rho_{\beta_2} $ & $u $ & $v $ &  $p$ & $z_1$ & $\lambda_1$ & $\lambda_2$ \Tstrut\Bstrut\\
    & $\left[\SI{}{\kilo\gram\per\cubic\meter}\right] $ & $\left[\SI{}{\kilo\gram\per\cubic\meter}\right]$ & $\left[\SI{}{\meter\per\second}\right]$ & $\left[\SI{}{\meter\per\second}\right]$ & $\left[\SI{}{\pascal}\right]$ & $\left[ - \right]$ & $\left[ - \right]$ & $\left[ - \right]$   \Tstrut\Bstrut\\[1mm]   \hline\hline
UF-TATB &  1800  & 1905 & 0 & 0 & $\SI{3.146e10}{}$ & $1-\SI{1e-6}{}$ & 0 & 1 \Tstrut\Bstrut\\ \hline
LX-17  & 1800  & 1905 & 0 & 0 & $\SI{1.0e5}{}$ & $\SI{1e-6}{}$ & 0 & 1 \Tstrut\Bstrut\\ \hline
\end{tabular} }
\vspace{2mm}
\caption{Initial conditions for the axisymmetric hockey puck configuration.}
\label{Tab:IC_HockeyPuck}
\end{table}

\subsubsection{Four-EoS rate stick configuration}

\begin{figure}[t]
  \centering
    \includegraphics[height=0.295\textwidth]{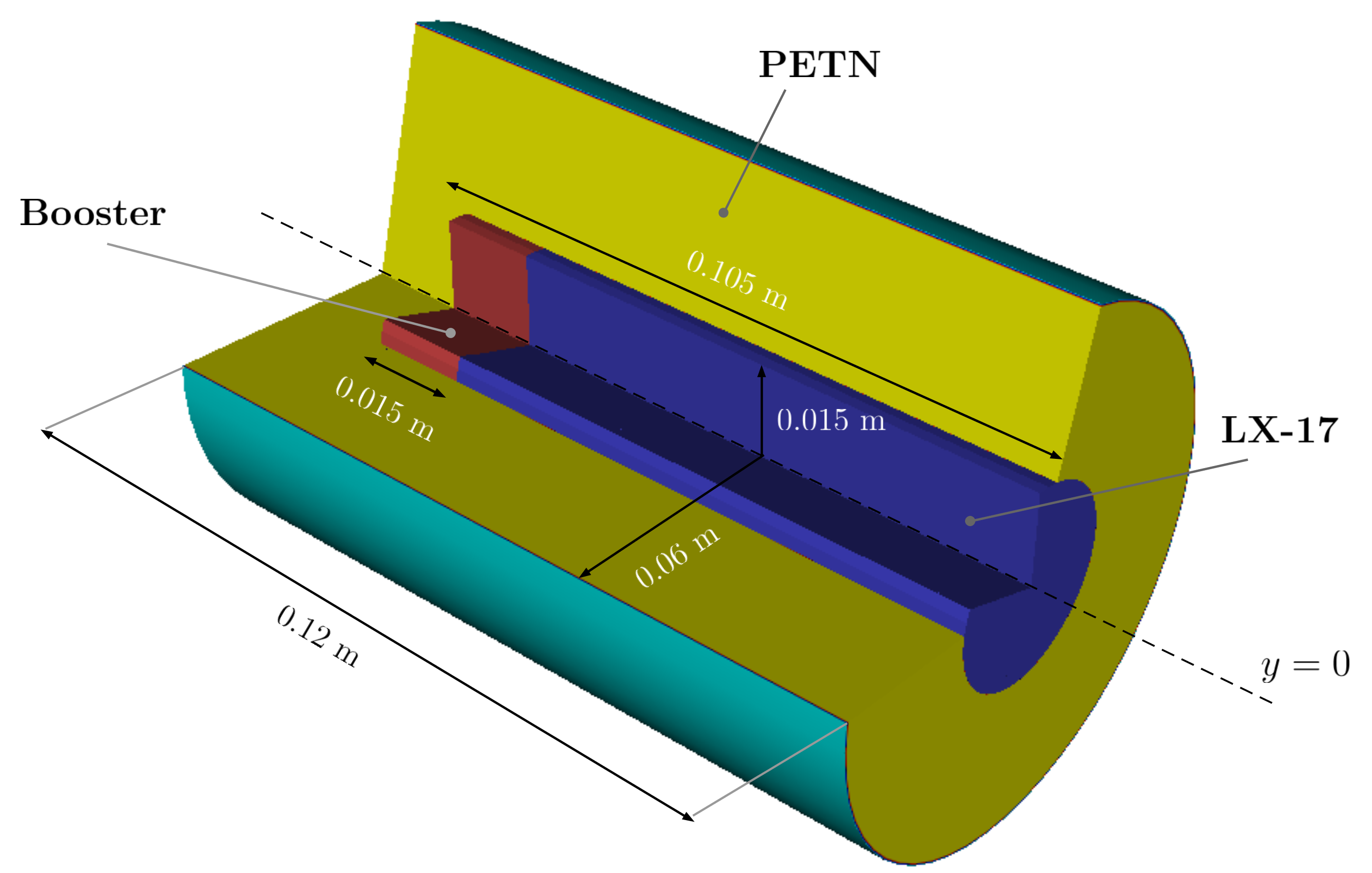} 
  \caption[Schematic illustration of the computational setup for the four-EoS rate stick problem.]{Schematic illustration of the computational setup for the four-EoS rate stick problem. The configuration is cylindrically symmetric with respect to the axis $y=0$.}
\label{Fig:4EOS_RateStick_Scheme}
\end{figure}

Next, we consider a rate stick configuration which is a modification of that studied by Michael and Nikiforakis in \cite{MINI16}, whereby the inert confiner is replaced by a second reactive material (PETN). Fig.~\ref{Fig:4EOS_RateStick_Scheme} shows the geometrical setup for this test; since the configuration is cylindrically symmetric with respect to the centreline of the rate stick, $y=0$, the problem can be solved in a two-dimensional domain with the addition of geometric source terms. The equations of state (JWL) and reaction rate law (Ignition and Growth) parameters for both explosives (LX-17 and PETN) are given in Tables \ref{Tab:EOS} and \ref{Tab:RateLaw}. The computational domain is the rectangle $[\SI{0}{}, \SI{0.12}{\meter}]\times[\SI{-0.06}{}, \SI{0}{\meter}]$, discretised with a base grid of $ 400 \times 200$ cells. Two levels of refinement, each with a refinement factor $\times 4$, are used to increase the resolution around the detonation waves in addition to continuous refinement at material interfaces. This results in an effective resolution of $\Delta x = \Delta y = \SI{18.75}{\micro \meter}$. A reflecting boundary condition is prescribed along the axis of symmetry ($y=0$) in order to account for the cylindrical nature of the problem. At all other sides, the boundary conditions are taken to be transmissive. The two explosive charges are initiated by means of a highly pressurised region located within the LX-17 at $(x,y) \in \left[ \SI{0.015}{\meter}, \SI{0.03}{\meter} \right] \times \left[ \SI{-0.015}{\meter}, \SI{0}{\meter}\right]$. The initial conditions are given in Table \ref{Tab:IC_4EOSRateStick}.
\begin{table}
  \vspace{1.5mm}
  \centering
	\renewcommand{\arraystretch}{1.25} 	
	\resizebox{0.485\textwidth}{!}{%
	\begin{tabular}{ c | c c c c c c c c} 
    \hline
    \multirow{2}{*}{Region} & $ \rho_{\alpha_1} = \rho_{\beta_1}$ & $\rho_{\alpha_2} = \rho_{\beta_2} $ & $u $ & $v $ &  $p$ & $z_1$ & $\lambda_1$ & $\lambda_2$ \Tstrut\Bstrut\\
    & $\left[\SI{}{\kilo\gram\per\cubic\meter}\right] $ & $\left[\SI{}{\kilo\gram\per\cubic\meter}\right]$ & $\left[\SI{}{\meter\per\second}\right]$ & $\left[\SI{}{\meter\per\second}\right]$ & $\left[\SI{}{\pascal}\right]$ & $\left[ - \right]$ & $\left[ - \right]$ & $\left[ - \right]$   \Tstrut\Bstrut\\[1mm]   \hline\hline
Booster &  1778  & 1905 & 0 & 0 & $\SI{3.37e10}{}$ & $\SI{1e-6}{}$ & 1 & 1 \Tstrut\Bstrut\\ \hline
LX-17 & 1778  & 1905 & 0 & 0 & $\SI{1.1e8}{}$ & $\SI{1e-6}{}$ & 1 & 1 \Tstrut\Bstrut\\ \hline
PETN  & 1778 & 1905 & 0 & 0 & $\SI{1.1e8}{}$ & $1-\SI{1e-6}{}$ & 1 & 1\Tstrut\Bstrut\\ \hline
\end{tabular} }
\vspace{2mm}
\caption{Initial conditions for the four-EoS cylindrical rate stick problem.}
\label{Tab:IC_4EOSRateStick}
\end{table}
\begin{figure}
    \centering
    {{a)}}
    \quad \hspace{0.4cm}
    \subfloat{\includegraphics[width=0.32\hsize]{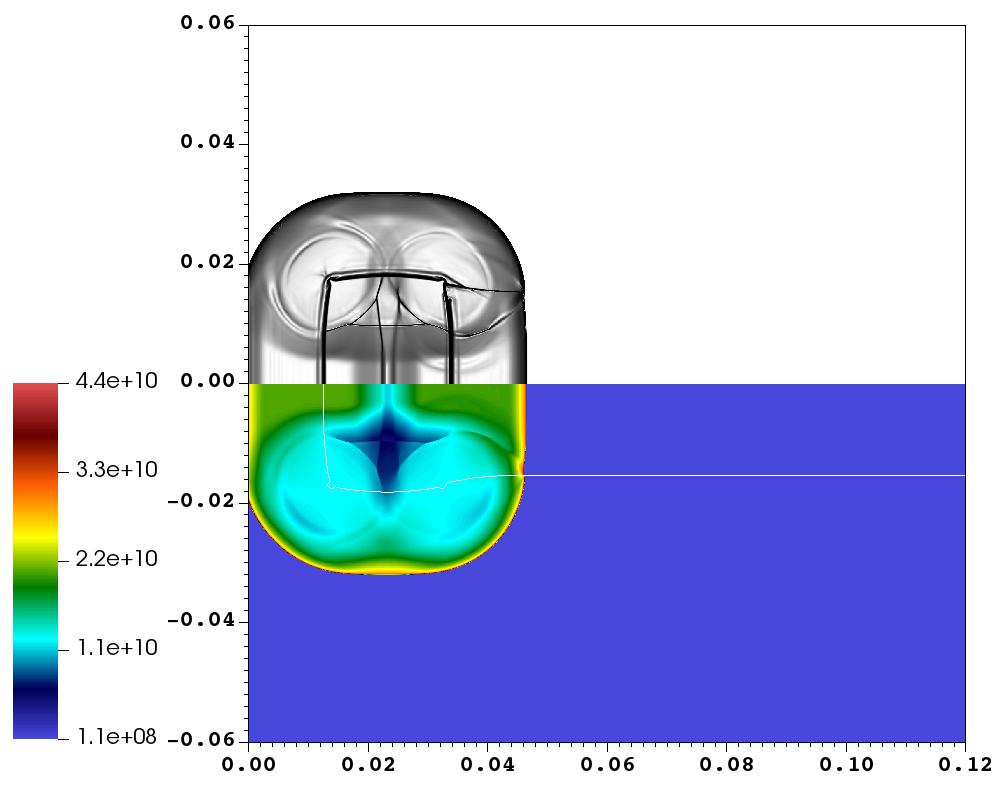}} \quad \hspace{1.8cm}
     \subfloat{\includegraphics[height=0.27\hsize]{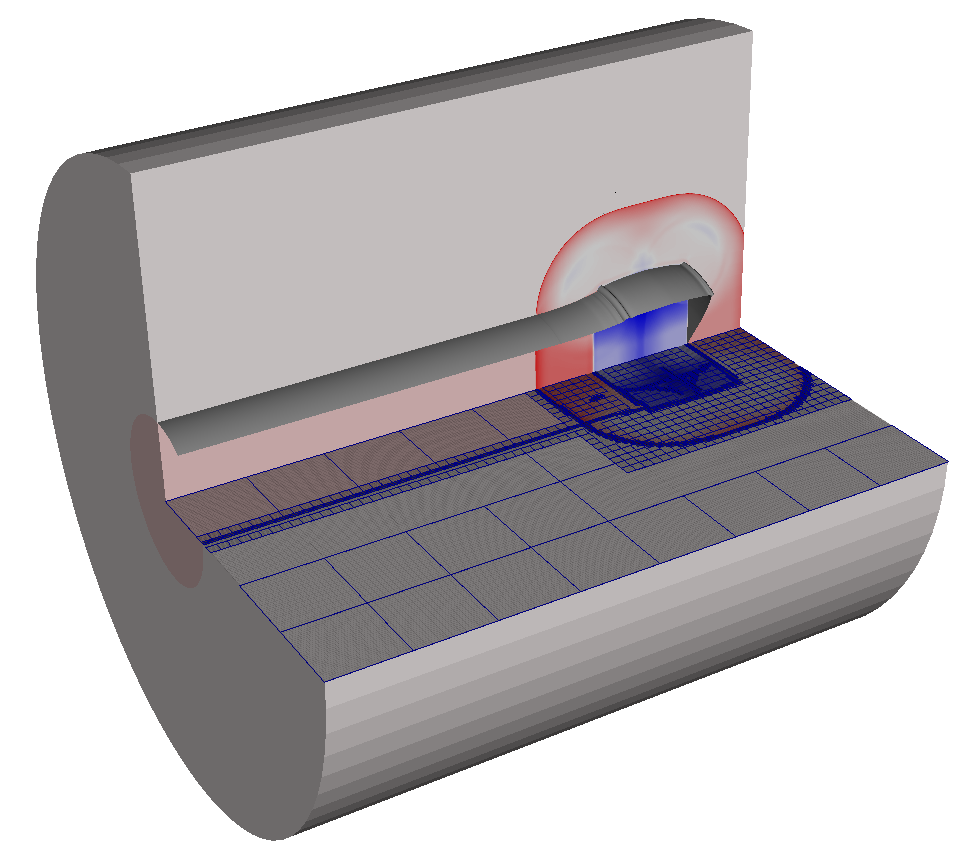}} \\[1.25ex]%
    {{b)}}
    \quad \hspace{0.4cm}
    \subfloat{\includegraphics[width=0.32\hsize]{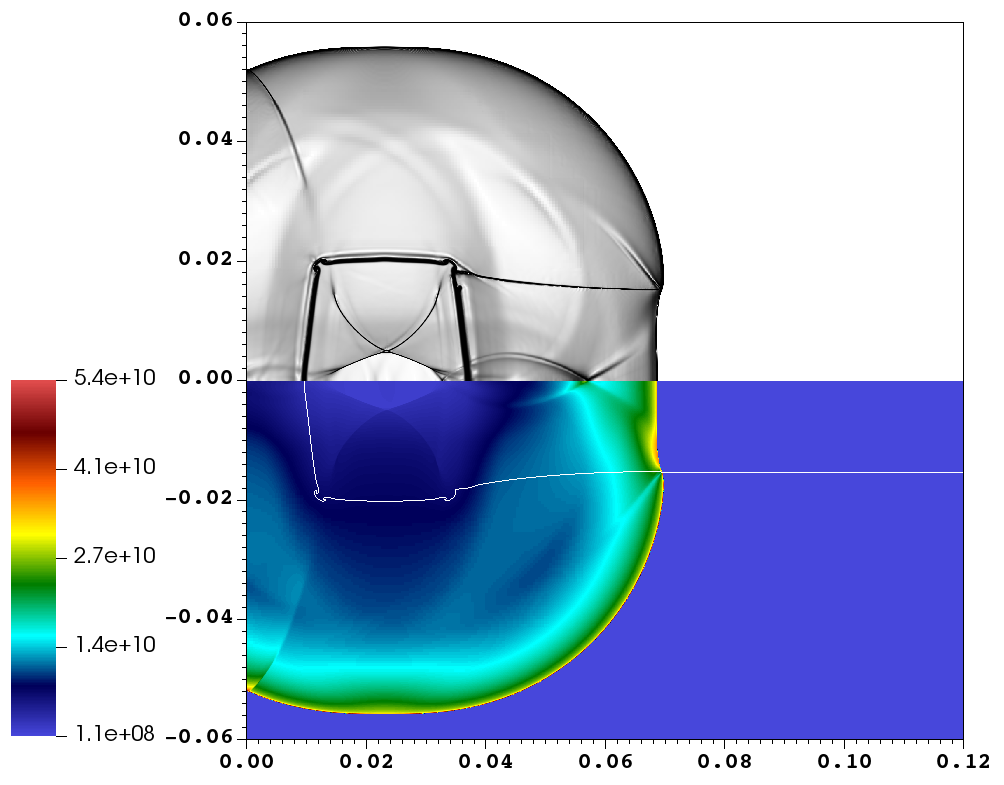}} \quad \hspace{1.8cm}
     \subfloat{\includegraphics[height=0.28\hsize]{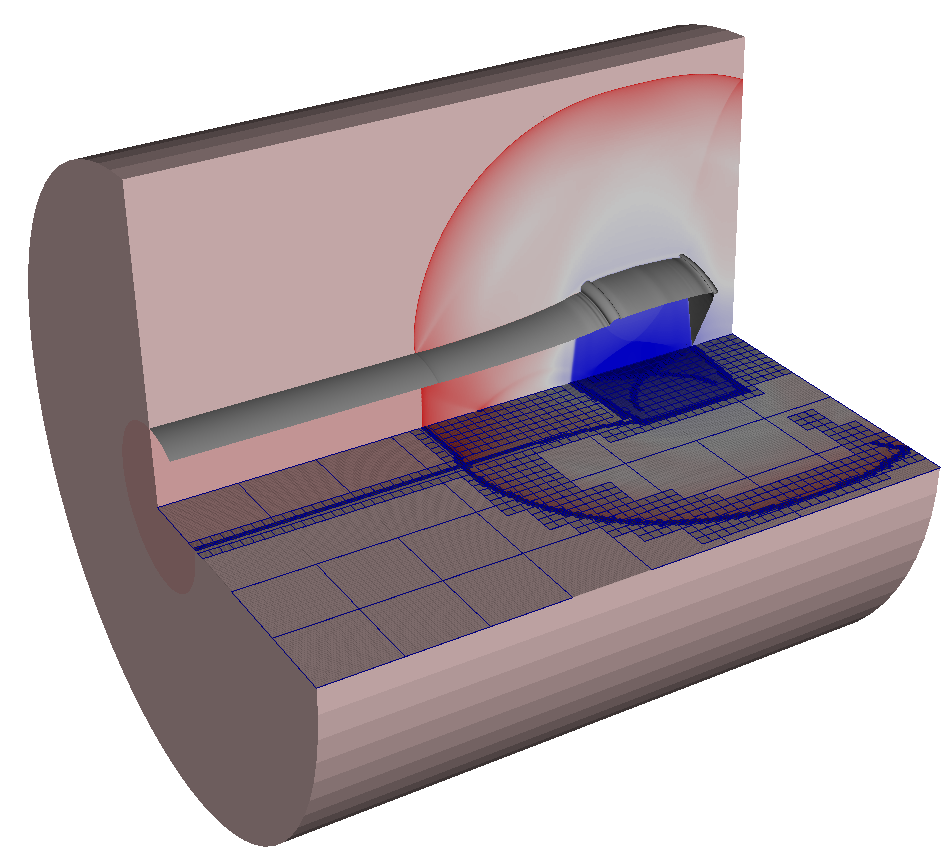}} \\[1.25ex]%
  {{c)}}
    \quad \hspace{0.4cm}
  \subfloat{\includegraphics[width=0.32\hsize]{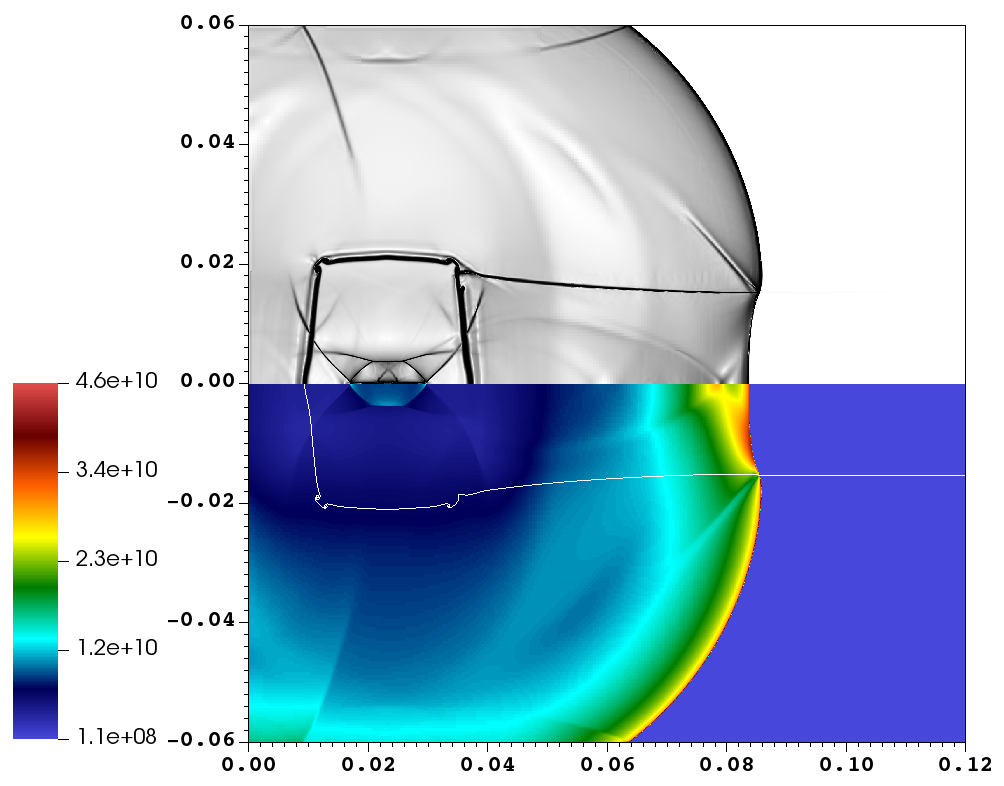}} \quad \hspace{1.8cm}
     \subfloat{\includegraphics[height=0.28\hsize]{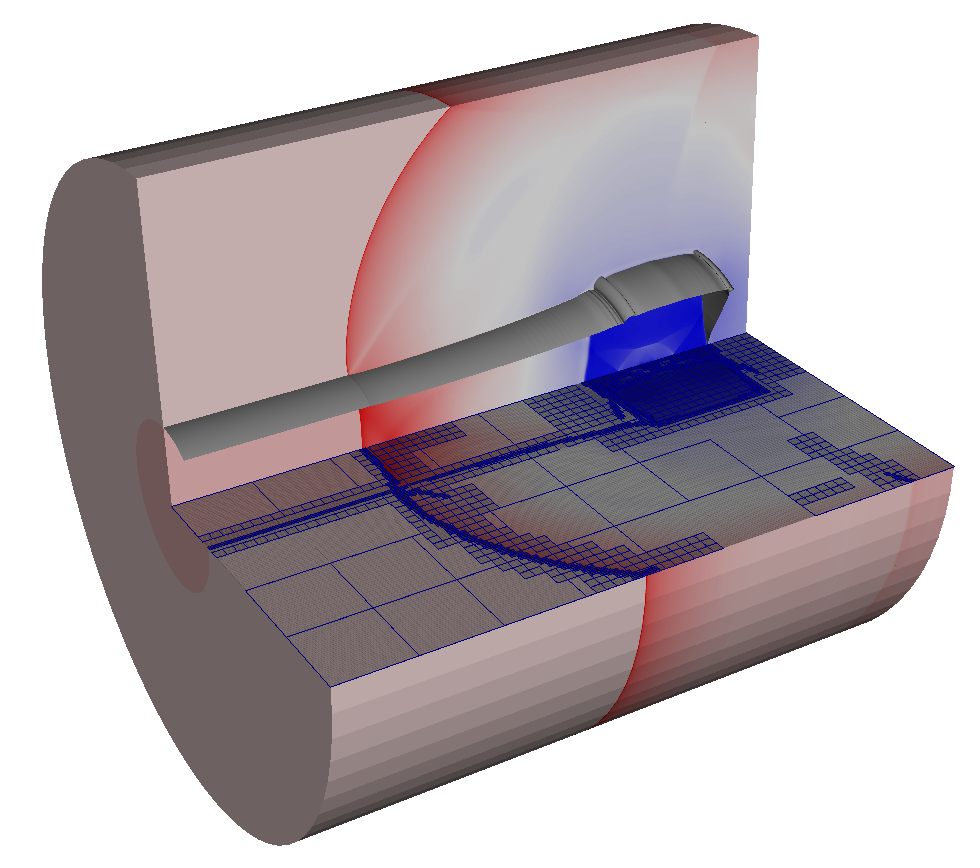}}
  \caption{Numerical solutions for the four-EoS rate stick problem at times a) $t=\SI{2}{\micro\second}$, b) $t=\SI{5}{\micro\second}$ and c) $t=\SI{7}{\micro\second}$. The figure illustrates the pressure field and density-based mock Schlieren plot (left) as well as the three-dimensional density field and the distribution of AMR grids (right). The two-explosive interface, defined as the contour $z_1=0.5$, is also shown and is identified by the white line superimposed to the pressure plot. Note that the figures of the right column have their orientation flipped to allow for better visualisation angles.}
  \label{Fig:Res4EosRatestick}
\end{figure}

Fig.~\ref{Fig:Res4EosRatestick} shows the pressure field and density-based mock Schlieren plot (left) as well as the three-dimensional density field and the AMR grids (right) at times $t=\SI{2}{\micro\second}$, $\SI{5}{\micro\second}$ and $\SI{7}{\micro\second}$. It is observed that the rapid expansion of the booster state leads to the prompt formation of two detonation waves: one propagating axially through the LX-17 stick and the other travelling radially outwards in PETN. These two waves interact at the two-explosive interface, where highly complicated refraction phenomena take place. In particular, it is seen that at the material interface the detonation wave in PETN travels ahead of the detonation front in the LX-17 stick, delivering energy to (and thus driving) this axially moving front of detonation, which attains a locally concave profile. This complex interaction occurring in the vicinity of the two-explosive interface also causes the local weakening of both detonation waves. Furthermore, as a result of the pressure imbalance produced in this region, the material interface is forced to expand outwards (see the white line in Fig.~\ref{Fig:Res4EosRatestick} indicating the LX-17/PETN interface), but this deformation is not as pronounced as it is in the three-EoS rate stick problem studied by Michael and Nikiforakis \cite{MINI16}, where an inert confiner was used. By the end of the simulation both detonations have approached steady state. In particular, the detonation wave travelling axially through the LX-17 stick attains a velocity of $D\sim0.987 D_{CJ}$ ($D_{CJ} = \SI{7679.9473}{\meter \second^{-1}}$) along the axis of symmetry, a value very similar to that obtained in the three-EoS rate stick problem. It is important to note that the simulation is free from numerical oscillations at the two-explosive interface as this interface is handled by the two-phase part of the mathematical model. Also, no robustness issues associated with two-dimensional root-finding are observed at grid points where the interaction between the two detonation waves takes place.  

\subsubsection{Single high explosive bead in liquid reactive nitromethane}
%
\begin{figure}[t]
  \centering
    \includegraphics[width=0.485\textwidth]{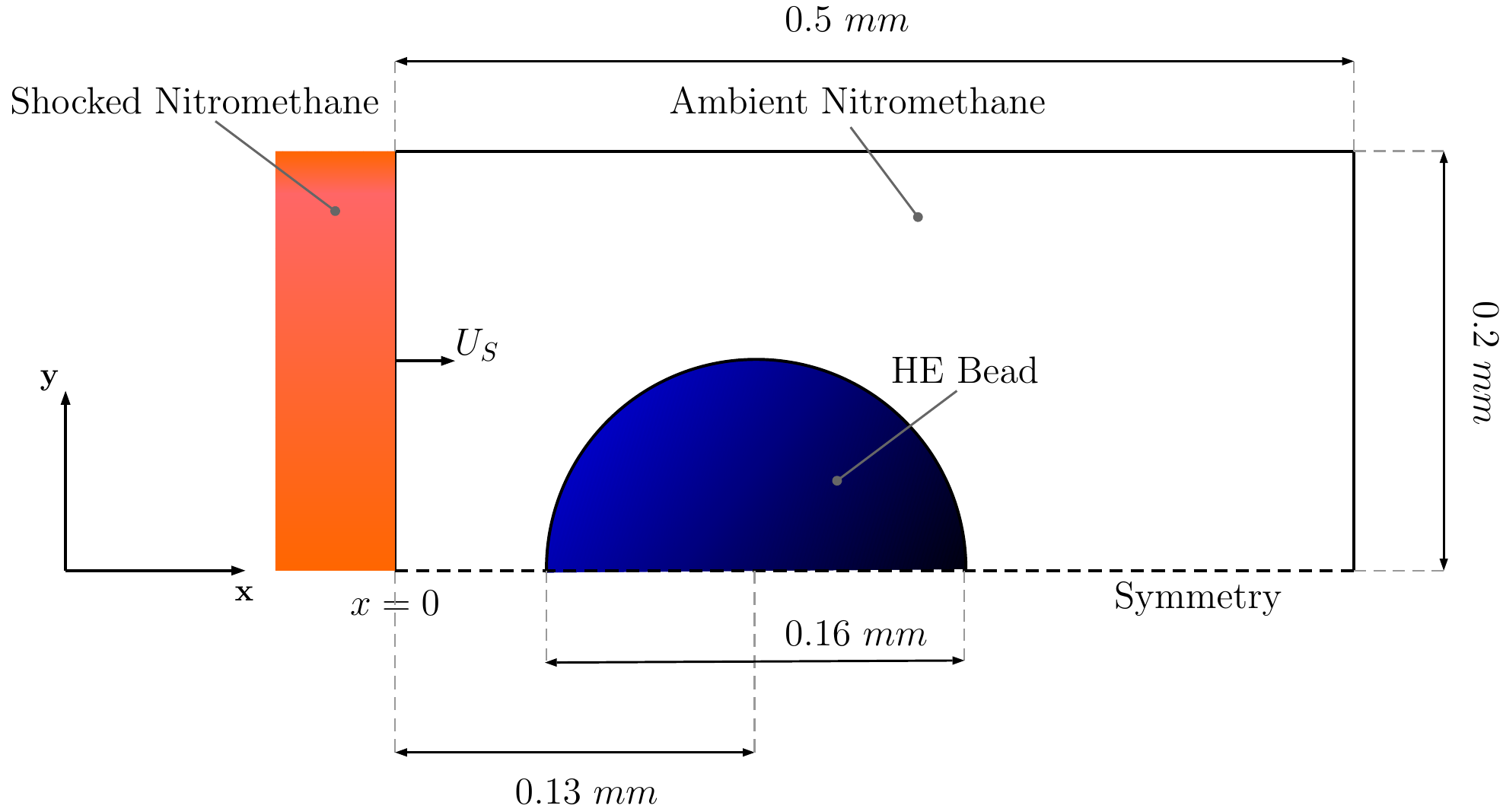} 
  \caption{Schematic illustration of the computational setup for an isolated high explosive bead interacting with a $\SI{10.98}{\giga\pascal}$ shock wave in liquid reactive nitromethane. At the initial time the shock is located at $x=\SI{0}{\milli\meter}$ and travels from left to right. The blue shaded area represents the LX-17 bead which has an initial radius of $ R = \SI{0.08}{\milli\meter}$.}
\label{Fig:Bead_schematic}
\end{figure}
\begin{table}[b!]
  \centering
	\renewcommand{\arraystretch}{1.25} 	
	\resizebox{0.485\textwidth}{!}{%
	\begin{tabular}{ c | c c c c c c c c} 
    \hline
     \multirow{2}{*}{Region} & $ \rho_1 $ & $\rho_2 $ & $u$ & $v $ &  $p $ & $z_1$ & $\lambda_1$ & $\lambda_2$ \Tstrut\Bstrut\\
    & $\left[\SI{}{\kilo\gram\per\cubic\meter}\right] $ & $\left[\SI{}{\kilo\gram\per\cubic\meter}\right]$ & $\left[\SI{}{\meter\per\second}\right]$ & $\left[\SI{}{\meter\per\second}\right]$ & $\left[\SI{}{\pascal}\right]$ & $\left[ - \right]$ & $\left[ - \right]$ & $\left[ - \right]$\Tstrut\Bstrut\\[1mm]   \hline\hline
Shocked Nitr. &  1905  & 1934 & 2000 & 0 & $\SI{10.98e9}{}$ & $\SI{1e-6}{}$ & 1 & 1 \Tstrut\Bstrut\\ \hline
Ambient Nitr. & 1905  & 1134 & 0 & 0 & $\SI{1e5}{}$ & $\SI{1e-6}{}$ & 1 & 1 \Tstrut\Bstrut\\ \hline
LX-17 bead & 1905 & 1134 & 0 & 0 & $\SI{1e5}{}$ & $1-\SI{1e-6}{}$ & 1 & 1\Tstrut\Bstrut\\ \hline
\end{tabular} }
\vspace{2mm}
\caption{Initial conditions for an isolated high explosive bead interacting with a $\SI{10.98}{\giga\pascal}$ shock wave in liquid reactive nitromethane.}
\label{Tab:IC_BEAD}
\end{table}

An important application of this work is the study of the sensitising character of reactive heterogeneities added into a liquid explosive matrix. To this end, we simulate an isolated high explosive (HE) bead embedded in liquid nitromethane, both igniting under the influence of an incident $\SI{10.98}{\giga\pascal}$ shock wave. This test takes inspiration from the numerical experiments carried out by Michael and Nikiforakis \cite{LouisaSolid}, where the effect of inert solid particles embedded in liquid nitromethane was assessed. The aim here is to study the influence of a single reactive particle on the ignition time of nitromethane. 

The computational setup is schematically illustrated in Fig.~\ref{Fig:Bead_schematic}. At the initial time the shock is located at $x=0$ and travels from left to right. The HE bead is centred at $(x,y) = \left(\SI{0.13}{\milli\meter}, \SI{0}{\milli\meter}\right)$, with an initial radius of $\SI{0.08}{\milli\meter}$. As the bead is symmetric about the horizontal axis only the upper part of the domain is simulated, while the lower part is obtained by symmetry applying an artificial reflective boundary condition at $y=\SI{0}{\milli\meter}$. All other domain boundary conditions are prescribed to be transmissive using simple zeroth-order extrapolation for all flow variables. The computational domain is the rectangle $[\SI{0}{\milli\meter},\SI{0.5}{\milli\meter}] \times [\SI{0}{\milli\meter}, \SI{0.2}{\milli\meter}]$, discretised with a base grid resolution of $320 \times 128$ cells. Two levels of refinement with refinement factors $\times 4$ and $\times 2$ are used, yielding the same effective resolution as in the shock-induced void collapse simulated by Michael and Nikiforakis \cite{Michael_Nikiforakis:2019_b}. 

The initial conditions for this test problem are given in Table \ref{Tab:IC_BEAD}. Nitromethane is modelled by the Cochran-Chan equation of state and a single-step Arrhenius reaction rate law, with parameters as given in Tables \ref{Tab:EOS} and \ref{Tab:RateLaw}. The HE bead is composed of LX-17, whose reactants and products are modelled using two distinct JWL equations of state with parameters as given in Table \ref{Tab:EOS}. The ignition term of the $I\&G$ model describing the reaction process in the LX-17 bead has been slightly modified so that the material is more easily and rapidly ignited by the passage of a shock, akin to a theoretically pre-sensitised explosive. This is done to demonstrate the effect of a fully developed reaction wave front propagating through the bead on the sensitising mechanism of nitromethane and thus help us gain a better understanding of the physics behind the sensitising mechanism. In particular, we set $a = 0$ and $\Phi_{ig,max} = 0.1$. The remaining reaction rate parameters are taken as in Table \ref{Tab:RateLaw}. Note that for this test the nitromethane products of reaction are not modelled explicitly and therefore a reduced version (requiring only one-dimensional root-finding) of our extended model is effectively used to simulate the scenario. 

\begin{figure}[t]
    \centering
    a.r) \quad \hspace*{-0.45cm}
    \subfloat{\includegraphics[width=0.29\hsize]{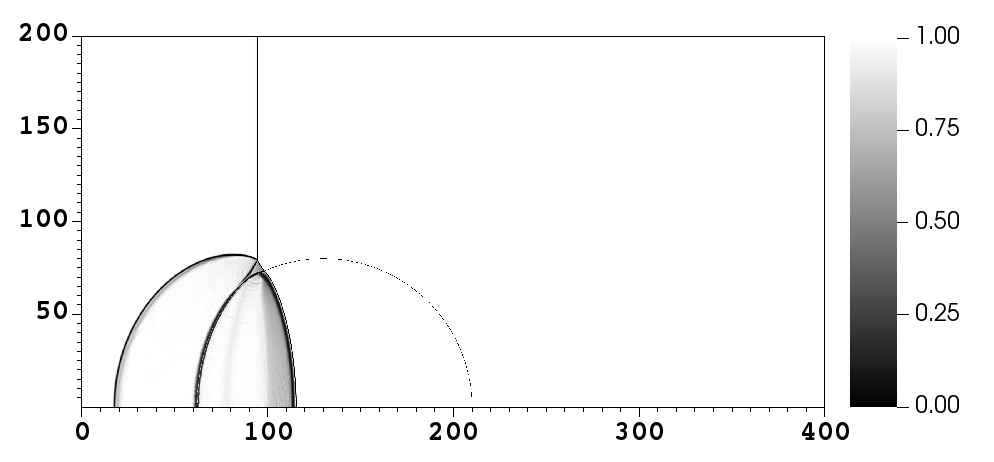}} \quad
    \subfloat{\includegraphics[width=0.29\hsize]{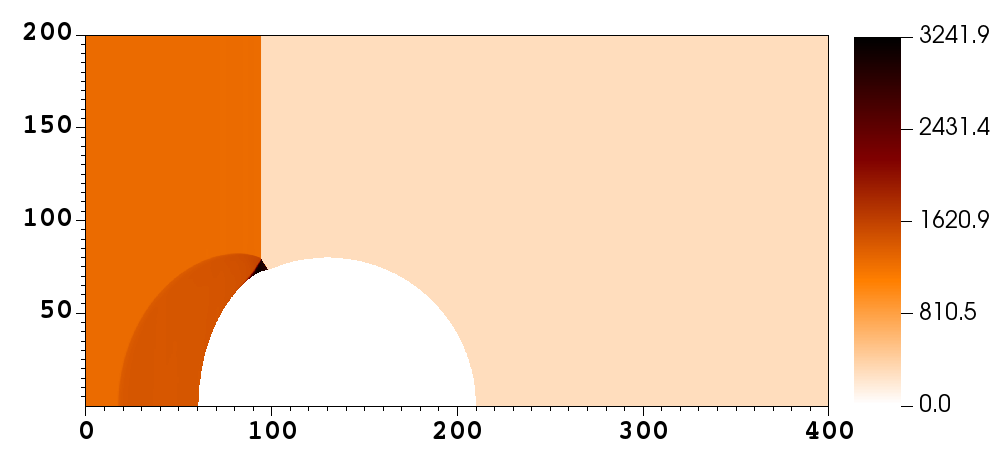}} \qquad
     \subfloat{\includegraphics[width=0.29\hsize]{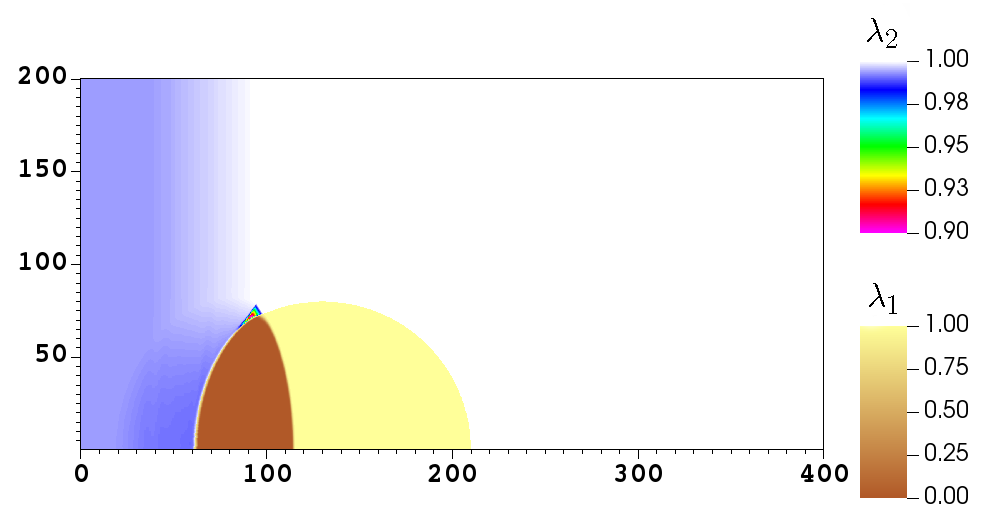}} \\[1ex]%
      a.nr) \quad \hspace*{-0.45cm}
    \subfloat{\includegraphics[width=0.29\hsize]{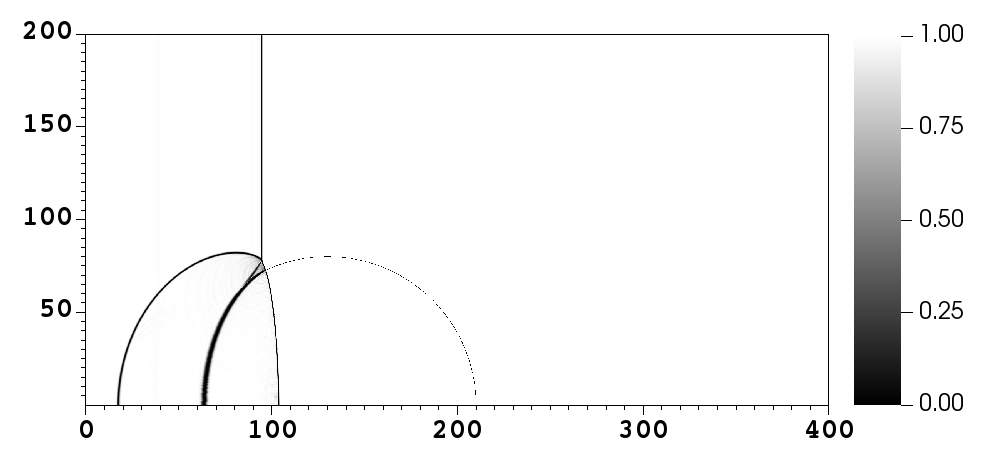}} \quad
    \subfloat{\includegraphics[width=0.29\hsize]{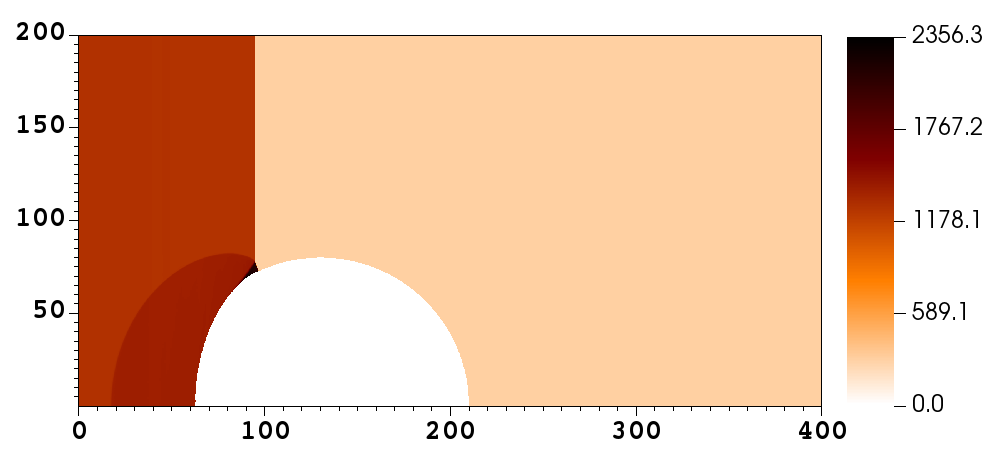}} \qquad
     \subfloat{\includegraphics[width=0.29\hsize]{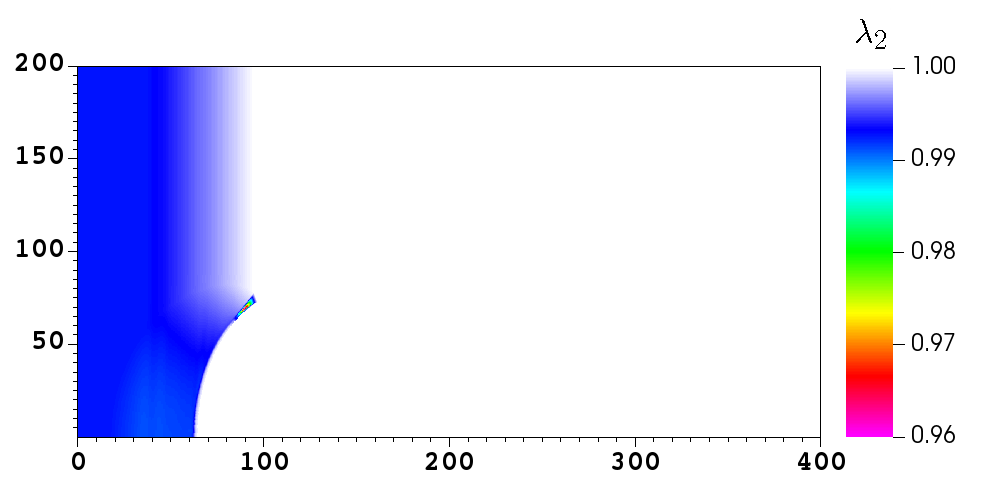}} \\[1ex]%
  b.r) \quad \hspace*{-0.45cm} \subfloat{\includegraphics[width=0.29\hsize]{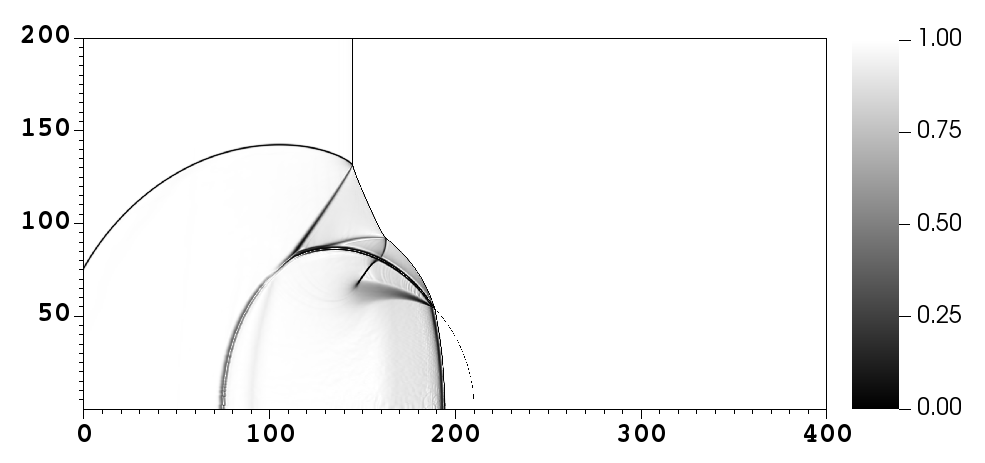}} \quad
    \subfloat{\includegraphics[width=0.29\hsize]{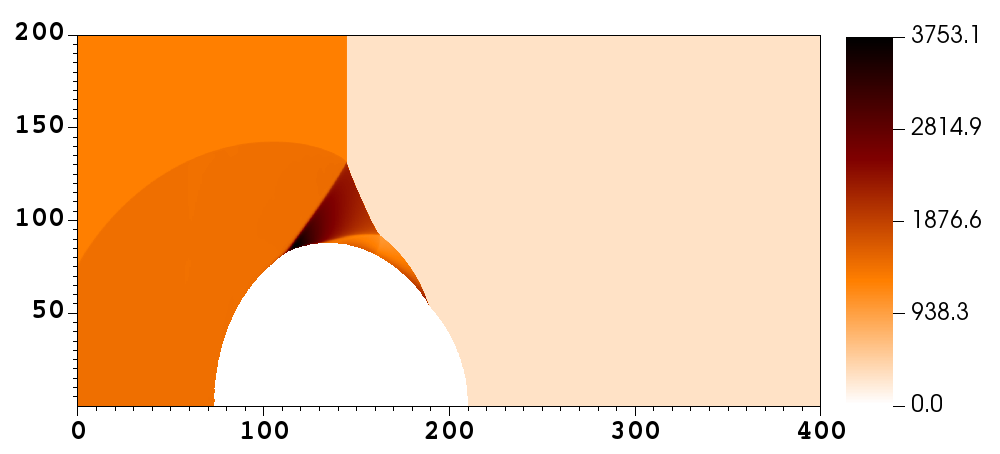}}\qquad
     \subfloat{\includegraphics[width=0.29\hsize]{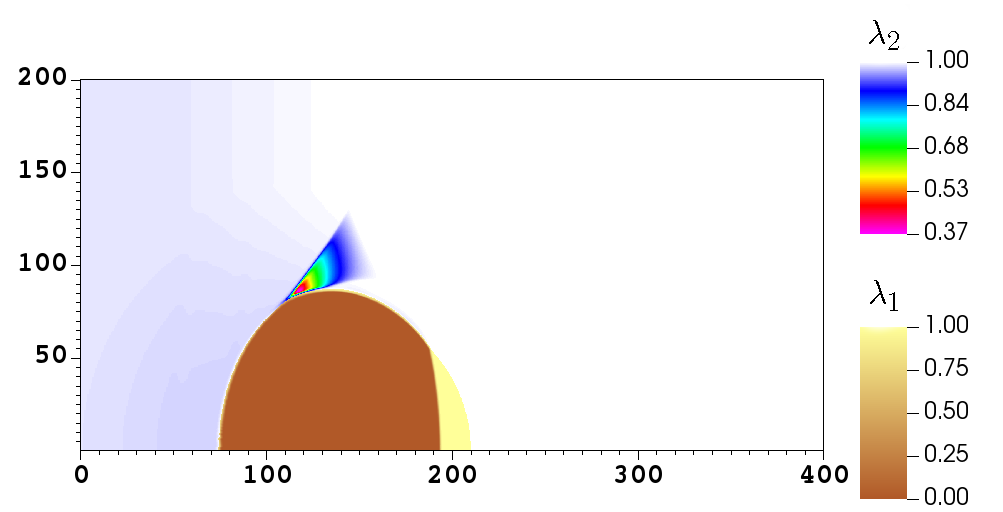}} \\[1ex]%
     b.nr) \quad \hspace*{-0.45cm} \subfloat{\includegraphics[width=0.29\hsize]{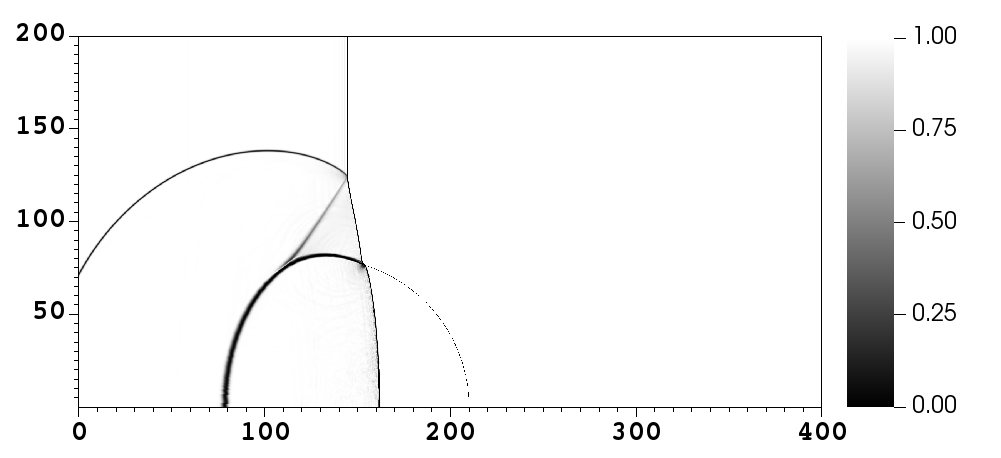}} \quad
    \subfloat{\includegraphics[width=0.29\hsize]{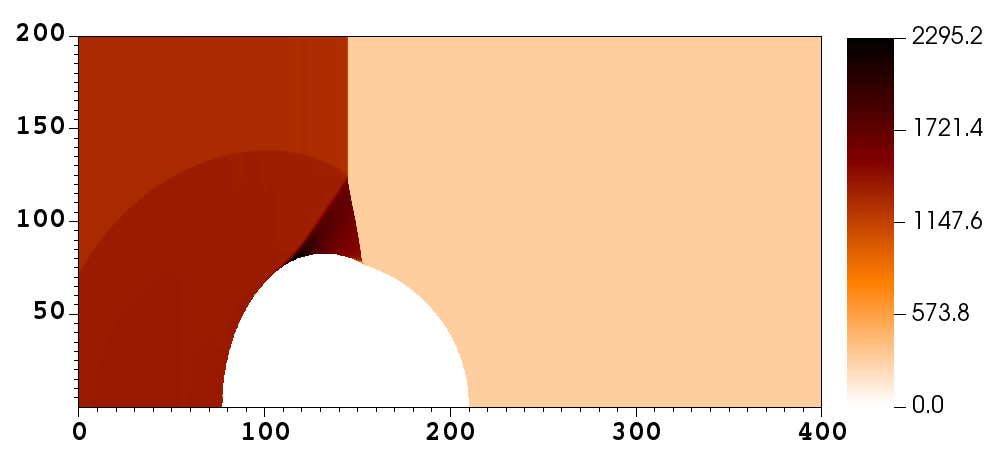}} \qquad
     \subfloat{\includegraphics[width=0.29\hsize]{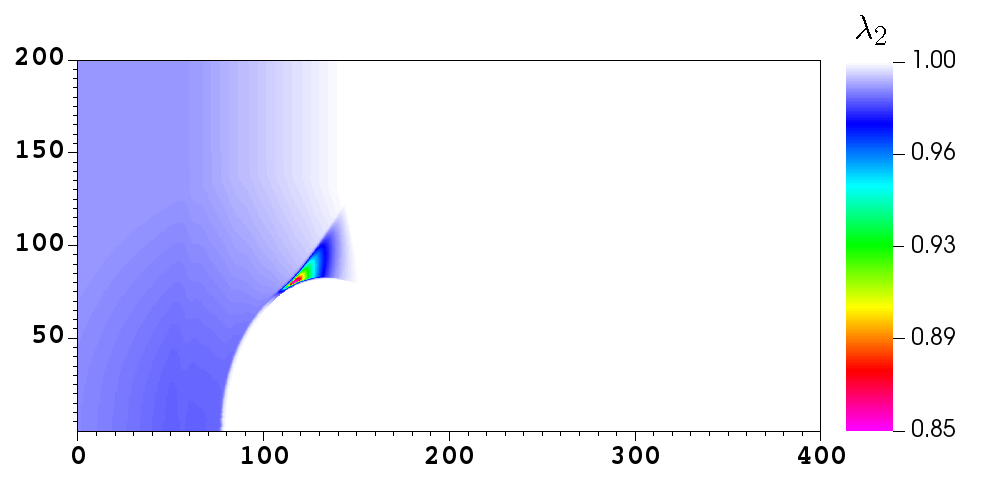}} \\[1ex]%
      c.r) \quad \hspace*{-0.45cm}
    \subfloat{\includegraphics[width=0.29\hsize]{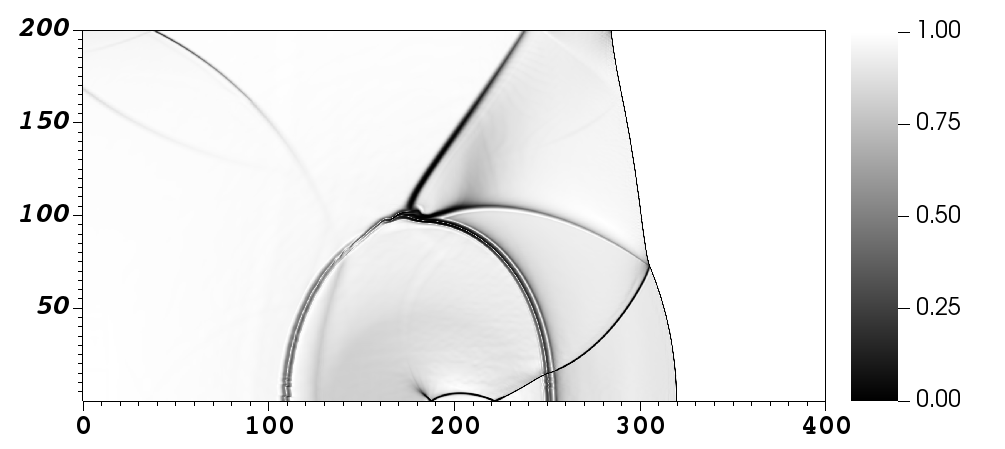}} \quad
    \subfloat{\includegraphics[width=0.29\hsize]{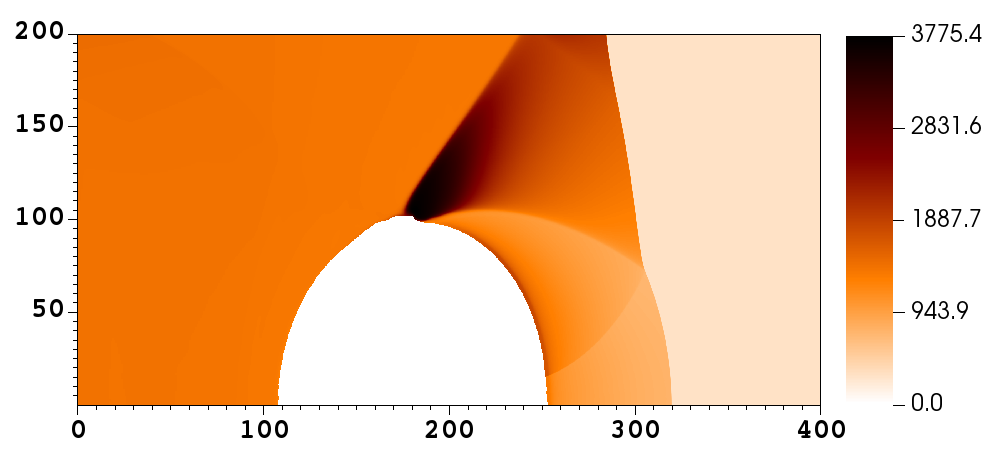}} \qquad
     \subfloat{\includegraphics[width=0.29\hsize]{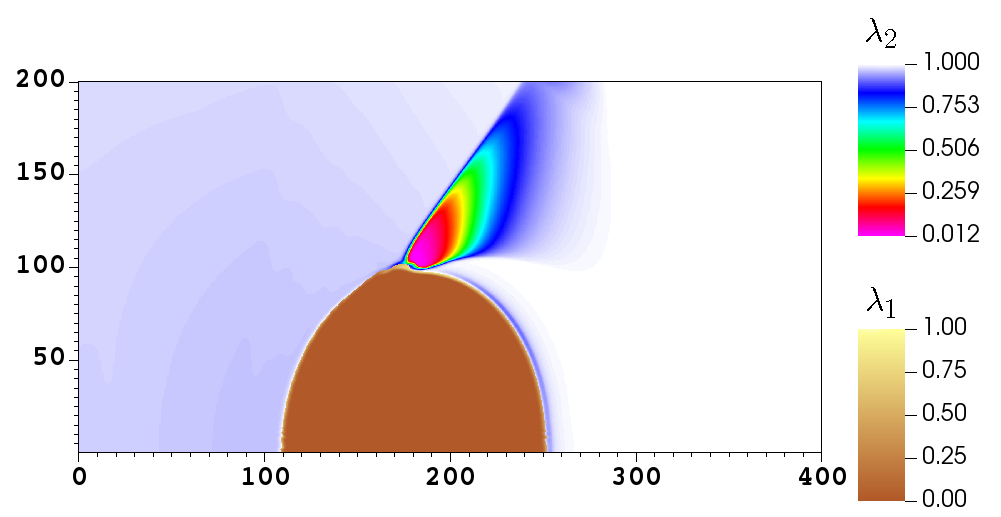}} \\[1ex]%
    c.nr) \quad \hspace*{-0.45cm}
   \subfloat{\includegraphics[width=0.29\hsize]{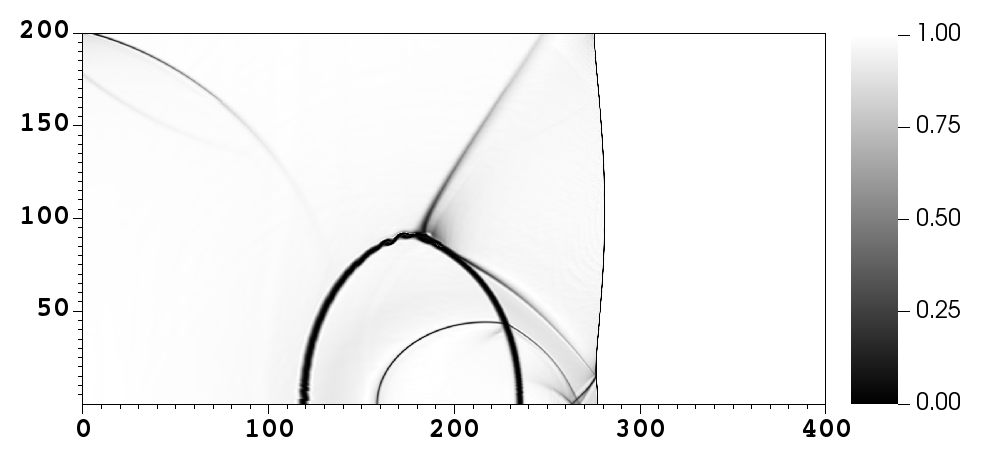}} \quad
    \subfloat{\includegraphics[width=0.29\hsize]{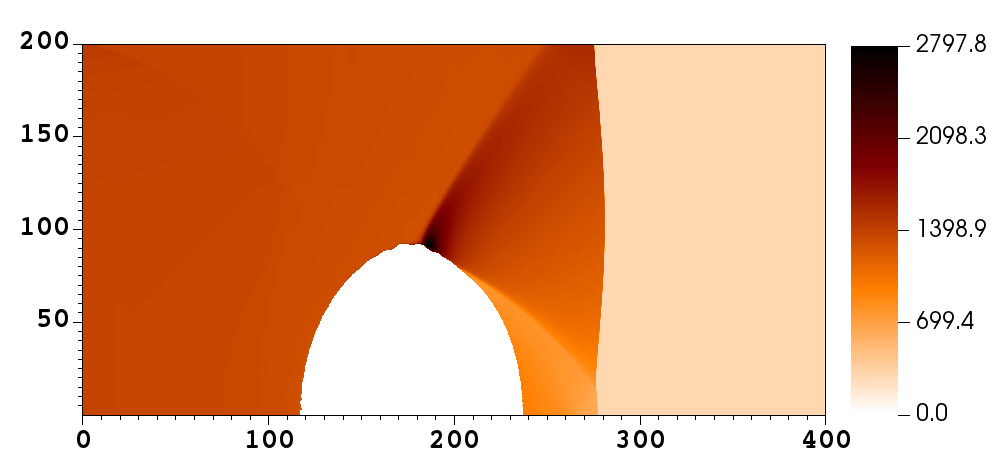}}\qquad
     \subfloat{\includegraphics[width=0.29\hsize]{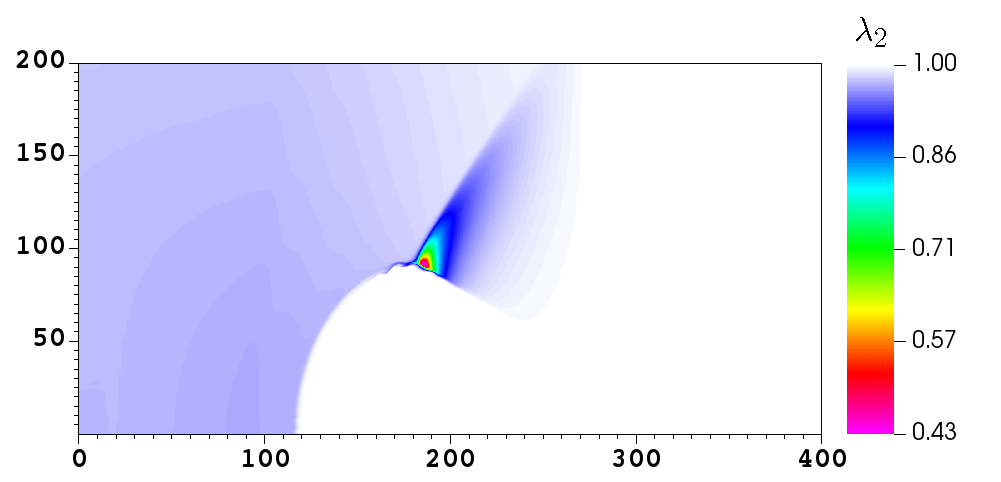}} \\[1ex]%
  \caption{Numerical results for an isolated HE bead interacting with a $\SI{10.98}{\giga\pascal}$ shock wave in liquid reactive nitromethane. Density mock-schlieren (left), nitromethane temperature (middle left), pressure (middle right) and reaction variables, $\lambda_1$ and $\lambda_2$, (right) plots are displayed at times a) $t= \SI{0.0185}{\micro\second} $, b) $\SI{0.029}{\micro\second}$ and c) $\SI{0.055}{\micro\second}$ . The interface between the bead and the nitromethane is also depicted and is identified by the white line superimposed to the pressure plots. For comparison, at each selected time we show the numerical results of a second simulation run with the same parameters but no reactive term in the bead --- referred to as the non-reacting case and denoted as ".nr" in the figure. The reacting case is instead denoted as ".r". The horizontal axis represents $x$ and the vertical axis represents $y$, both in $\SI{}{\micro\meter}$. Pressure is expressed in $\left[\SI{}{\pascal}\right]$ and temperature in $\left[\SI{}{\kelvin}\right]$. }
 \label{Fig:HEBead}
\end{figure}

Fig.~\ref{Fig:HEBead} illustrates the density mock-schlieren, pressure, nitromethane temperature and reaction variables ($\lambda_1$ and $\lambda_2$) plots at selected times. For the sake of comparison, at each selected time we also show the numerical results of a second simulation run with the same parameters but no reactive term in the bead --- hereinafter referred to as the non-reacting case. In this way we can compare the two cases to better understand the influence of the reactivity of the bead on the time to ignition and modes of ignition of nitromethane.


Initially, the shock wave propagates through the ambient nitromethane, compressing the material to $\SI{10.98}{\giga\pascal}$ and raising the temperature to $\sim \SI{1270}{\kelvin}$. The interaction of the incident shock with the high explosive bead generates two new shock waves, one propagating upstream and re-compressing the nitromethane and one travelling downstream into the bead. As the transmitted shock propagates through the reactive bead, the reaction process is triggered almost instantly and a reaction wave is formed; naturally, this reaction wave is not present in the non-reacting case. Due to the curvature of the HE particle, a transition from regular reflection to Mach reflection is then observed to occur at time $t \sim \SI{0.016}{\micro\second}$, generating a Mach stem at the top of the bead.

Interestingly, the energy released by the combustion of the bead is observed to increase the strength of the Mach wave, which heats the nitromethane to a temperature of about $\SI{3250}{\kelvin}$, approximately $\SI{900}{\kelvin}$ higher than that reached in the same region in the non-reacting case. This rapid rise in temperature leads to the formation of a hot spot locus with the consequent local ignition of nitromethane, as seen in the $\lambda_1-\lambda_2$ plot of Fig.~\ref{Fig:HEBead}a.r). Using the standard $10\%$--definition \cite{Michael_Nikiforakis:2019_b}, the ignition is observed to occur at $t \sim \SI{0.0185}{\micro\second}$ for the reacting case, while ignition occurs at a later time ($t\sim\SI{0.0265}{\micro\second}$) for the non-reacting case. As time progresses, the Mach stem region enlarges and the Mach stem triple point moves away from the bead, along the incident shock, producing a band of high temperatures, where most of the reaction is concentrated. At time $t=\SI{0.029}{\micro\second}$, this zone exhibits temperatures as high as $\sim \SI{3750}{\kelvin}$ in the reacting case, while these reach up to only $\sim \SI{2300}{\kelvin}$ in the non-reacting case. In the final stages of the simulation, multiple and complex reflection phenomena are visible inside the bead, which is deformed and accelerated downstream; more burning is observed in the reaction site, with $\lambda_2$ reaching a value of $0.012$ in the reacting case and a value of $0.43$ in the non-reacting case by the end of the simulation ($t=\SI{0.055}{\micro\second}$).

The main conclusion that can be drawn from these results is that the presence of an isolated solid particle strongly modifies the temperature field distribution, generating regions of locally high temperatures and leading to a further reduction of the time to ignition of nitromethane. It is thus clear that reactive particles added into an explosive matrix can significantly affect the ignition sensitivity of the material and may represent another effective means of controlling its performance. 
 
\section{Conclusions}

In this work we presented a mathematical formulation and the corresponding algorithm for its numerical solution for the interaction between two reactive materials in direct contact. To this end, the original system of nonlinear partial differential equations proposed by Michael and Nikiforakis \cite{MINI16} is extended to account for this case. A new numerical algorithm is presented for its numerical solution, which includes a robust root-finding procedure which takes into account physical as well as mathematical criteria in order to converge to the correct solution. The new model preserves the desirable features of the original MiNi16 algorithm, and it is suitable for the robust solution of problems involving military-grade and commercial explosives in direct contact, each having distinct equations of state for both the unreacted phase and the gaseous products. The formulation was validated against known solutions in the literature and the applicability of the model to scenarios involving two reactive media was presented.

\bibliographystyle{plain}
\bibliography{aipsamp}

\end{document}